\documentclass[aps,showpacs,preprintnumbers,amsmath, amssymb]{revtex4}

\usepackage{float}
\usepackage{graphics,epsfig}
\usepackage{graphicx}
\usepackage{dcolumn}
\usepackage{bm}

\begin{document}
\baselineskip=0.8 cm
\title{{\bf Notes on holographic superconductor models with the nonlinear electrodynamics}}

\author{Zixu Zhao$^{1,2}$, Qiyuan Pan$^{1,2}$\footnote{panqiyuan@126.com}, Songbai Chen$^{1,2}$\footnote{csb3752@hunnu.edu.cn} and Jiliang Jing$^{1,2}$\footnote{jljing@hunnu.edu.cn}}
\affiliation{$^{1}$Institute of Physics and Department of Physics,
Hunan Normal University, Changsha, Hunan 410081, China}
\affiliation{$^{2}$ Key Laboratory of Low Dimensional Quantum
Structures and Quantum Control of Ministry of Education, Hunan
Normal University, Changsha, Hunan 410081, China}

\vspace*{0.2cm}
\begin{abstract}
\baselineskip=0.6 cm
\begin{center}
{\bf Abstract}
\end{center}

We investigate systematically the effect of the nonlinear correction to the usual Maxwell electrodynamics on the holographic dual models in the backgrounds of AdS black hole and AdS soliton. Considering three types of typical nonlinear electrodynamics, we observe that in the black hole background the higher nonlinear electrodynamics correction makes the condensation harder to form and changes the expected relation in the gap frequency, which is similar to that caused by the curvature correction. However, in strong contrast to the influence of the curvature correction, we find that in the AdS soliton background the nonlinear electrodynamics correction will not affect the properties of the holographic superconductor and insulator phase transitions, which may be a quite general feature for the s-wave holographic superconductor/insulator system.

\end{abstract}


\pacs{11.25.Tq, 04.70.Bw, 74.20.-z}\maketitle
\newpage
\vspace*{0.2cm}

\section{Introduction}

In modern condensed matter physics, one of the unsolved mysteries is the core mechanism governing the high-temperature superconductor systems which are not described by the usual BCS theory. Interestingly, recent developments of the holography by applying that to the strongly coupled condensed matter
system, especially the construction of the holographic superconductor, might give some insights
into the pairing mechanism in the high $T_c$ superconductors \cite{HartnollJHEP12}. From the anti-de Sitter/conformal field theory (AdS/CFT) correspondence, which states that a weak coupling gravity theory in a $d$-dimensional AdS spacetime can be related to a strong CFT on the $(d-1)$-dimensional boundary \cite{Maldacena,Gubser1998,Witten}, it was suggested that the instability of the bulk black hole corresponds to a second order phase transition from normal state to superconducting state which brings the spontaneous U(1) symmetry breaking \cite{GubserPRD78}. The authors of Ref. \cite{HartnollPRL101} introduced the first holographic superconductor model and observed that the properties of a ($2+1$)-dimensional superconductor can indeed be reproduced in this simple model. Since then, a large number of the holographic dual models have been constructed and some interesting behaviors have been found, for reviews, see Refs. \cite{HartnollRev,HerzogRev,HorowitzRev} and references therein.

In most cases, the studies on the holographic superconductors focus on the Einstein-Maxwell theory coupled to a charged scalar field. In the light of AdS/CFT correspondence, the higher derivative corrections to either gravitational or electromagnetic action in AdS space are expected to modify the dynamics of the strongly coupled dual theory. Motivated by the application of the Mermin-Wagner theorem to the holographic superconductors, there was an interesting study of the effect of a particularly gravitational correction, i.e., the curvature correction on the $(3+1)$-dimensional superconductor \cite{Gregory}. It was found that the higher curvature correction makes the condensation harder to form and changes the ratio $\omega/T_c\approx8$ in the gap frequency \cite{Gregory,Pan-Wang,Ge-Wang}. Extending the gravitational construction to include a Ricci flat AdS soliton in Gauss-Bonnet gravity \cite{CaiKimWang}, we observed that the higher curvature corrections make it harder for the superconductor/insulator phase transition to be triggered \cite{Pan-Wang,Pan-Jing-Wang-Soliton}. Besides the correction in gravity, it is of great interest to consider the higher derivative correction to the gauge field in order to understand the influences of the $1/N$ or $1/\lambda$ ($\lambda$ is the 't Hooft coupling) corrections on the holographic dual models. As a matter of fact, the nonlinear electrodynamics, which essentially implies the higher derivative corrections of the gauge field, carries more information than the usual Maxwell electrodynamics \cite{BornInfeld,Hoffmann,HeisenbergEuler,Oliveira,GibbonsRasheed} and has been a focus for these years since most physical systems are inherently nonlinear to some extent \cite{Anninos,Hendi,Miskovic,Shabad,Gurtug}. Considering the holographic superconductor models in the Born-Infeld electrodynamics, Jing and Chen found that the Born-Infeld coupling parameter will make it harder for the scalar condensation to form \cite{JS2010}. In the St$\ddot{u}$ckelberg mechanism, rich physics in the phase transition of the holographic superconductor with Born-Infeld electrodynamics in Gauss-Bonnet gravity has been observed \cite{JLQS2012}. Along this line, there have been accumulated interest to study various holographic dual models in the presence of various higher derivative (nonlinear) corrections to the usual Maxwell electrodynamics \cite{WuCKW,MaCW,JingJHEP,DMMPLA,PJWPRD,MomeniSL,LeeEPJC,SDSL2012,LPW2012,BGRL2012,DMRIJMPA,
JPCPLB,Roychowdhury,BGQX,ZPJ2012}.

More recently, Hendi introduced a new Born-Infeld-like nonlinear electrodynamics, i.e., Exponential form of nonlinear electrodynamics (ENE) in order to obtain the new charged BTZ black hole solutions in the Einstein-nonlinear electromagnetic field, motivated from obtaining a finite value for the self-energy of a pointlike charge \cite{HendiJHEP}. Compared with the Born-Infeld nonlinear electrodynamics (BINE) and Logarithmic form of nonlinear electrodynamics (LNE), the ENE has different effect on the electric potential and temperature for the obtained solutions. Thus, the motivation for completing this work is two-fold. On one level, it seems to be an interesting investigation to construct the holographic dual models with the ENE both in the backgrounds of AdS black hole and AdS soliton, which allows us to examine the effect of the ENE on the formation of the scalar hair and the relation connecting the gap frequency in conductivity with the critical temperature. On another more speculative level, it would be important to study systematically some typical nonlinear electrodynamics and see some general feature for the effects of the higher derivative corrections to the gauge field on the holographic dual models. We will consider a gauge field and the scalar field coupled via a generalized Lagrangian
\begin{eqnarray}\label{System}
S=\int d^{d}x\sqrt{-g}\left\{\frac{1}{16\pi
G}\left[R+\frac{(d-1)(d-2)}{L^2}\right]
+\left[\mathcal{L}(F^{2})-|\nabla\psi-iA\psi|^{2}-m^2|\psi|^2\right]\right\} \ ,
\end{eqnarray}
where $\mathcal{L}(F^{2})$ is the Lagrangian of three classes of Born-Infeld-like nonlinear electrodynamics
\begin{eqnarray}\label{NEF}
\mathcal{L}(F^{2})=\left\{
\begin{array}{rl}
\frac{1}{4\beta^2}\left(e^{-\beta^2F^{2}}-1\right)   \ , &  \quad {\rm ENE}\\
\frac{1}{\beta^2}\left(1-\sqrt{1+\frac{1}{2}\beta^2F^{2}}\right)   \ , &  \quad {\rm BINE} \\
-\frac{2}{\beta^2}\ln\left(1+\frac{1}{8}\beta^2F^{2}\right)   \ , &  \quad {\rm LNE}
\end{array}\right.
\end{eqnarray}
with the quadratic term $F^{2}=F_{\mu\nu}F^{\mu\nu}$. When the nonlinearity parameter $\beta\rightarrow0$, $\mathcal{L}(F^{2})$ obviously reduces to the standard Maxwell form $\mathcal{L}(F^{2})=-\frac{1}{4}F^{2}$. Note that the higher order terms in the parameter $\beta$ essentially correspond to the higher derivative corrections of the gauge fields \cite{Roychowdhury}. With the same value of $\beta$, we can discuss the difference in the three types of the holographic dual models with the nonlinear electrodynamics quantitatively. It should be noted that the horizon geometry of nonlinear charged black holes is close to the horizon of uncharged (Schwarzschild) black hole solution for very large values of $\beta$ \cite{HendiJHEP}, so in this case $\mathcal{L}(F^{2})$ can be neglected. In order to extract the main physics, in this work we will concentrate on the probe limit to avoid the complex computation.

The structure of this work is as follows. In Sec. II, we will investigate the holographic superconducting model with the ENE which has not been constructed as far as we know, and compare it with the BINE and LNE holographic superconductors. In Sec. III, we will extend the discussion to holographic superconductor/insulator transitions with these three kinds of typical nonlinear electrodynamics. We will summarize our results in the last section.

\section{Holographic superconducting models with the nonlinear electrodynamics}

In order to construct a superconductor dual to an AdS black hole configuration in the probe
limit, we consider the background of the $d$-dimensional planar
Schwarzschild-AdS black hole
\begin{eqnarray}\label{BH metric}
ds^2=-f(r)dt^{2}+\frac{dr^2}{f(r)}+r^{2}dx_{i}dx^{i},
\end{eqnarray}
where
\begin{eqnarray}
f(r)=\frac{r^2}{L^2}\left(1-\frac{r_{+}^{d-1}}{r^{d-1}}\right),
\end{eqnarray}
 $L$ is the AdS radius and $r_{+}$ is the radius of the event
horizon. The Hawking temperature of the black
hole, which will be interpreted as the temperature of the
CFT, can be expressed as
\begin{eqnarray}
T=\frac{(d-1)r_{+}}{4\pi L^2}.
\end{eqnarray}

Taking the ansatz of the matter fields as $\psi=|\psi|$, $A_{t}=\phi$ where $\psi$, $\phi$
are both real functions of $r$ only, we can get the equation of motion for the scalar field
$\psi$ from the action (\ref{System}) in the probe limit
\begin{eqnarray}
&&\psi^{\prime\prime}+\left(
\frac{d-2}{r}+\frac{f^\prime}{f}\right)\psi^\prime
+\left(\frac{\phi^2}{f^2}-\frac{m^2}{f}\right)\psi=0\,,
\label{BHPsi}
\end{eqnarray}
which is the same form for the three types of nonlinear electrodynamics (\ref{NEF}). For the gauge field $\phi$, however, we obtain the following equations of motion
\begin{eqnarray}
&&(1+4\beta^{2}\phi^{\prime 2})\phi^{\prime\prime}+\frac{d-2}{r}\phi^\prime-\frac{2\psi^{2}}{f}e^{-2\beta^{2}\phi^{\prime 2}}\phi=0~, \quad {\rm ENE} \nonumber\\
&&\phi^{\prime\prime}+\frac{d-2}{r}(1-\beta^{2}\phi^{\prime 2})\phi^\prime-\frac{2\psi^{2}}{f}(1-\beta^{2}\phi^{\prime 2})^{3/2}\phi=0~, \quad {\rm BINE} \nonumber\\ &&\left(1+\frac{1}{4}\beta^{2}\phi^{\prime 2}\right)\phi^{\prime\prime}+\frac{d-2}{r}\left(1-\frac{1}{4}\beta^{2}\phi^{\prime 2}\right)\phi^\prime-\frac{2\psi^{2}}{f}\left(1-\frac{1}{4}\beta^{2}\phi^{\prime 2}\right)^{2}\phi=0~, \quad {\rm LNE} \label{BHPhi}
\end{eqnarray}
Obviously, Eqs. (\ref{BHPsi}) and  (\ref{BHPhi})
reduce to the standard holographic superconductor models discussed
in \cite{HartnollPRL101,HartnollJHEP12,HorowitzPRD78} when
$\beta\rightarrow0$. At the event horizon $r=r_{+}$ of the black hole, the regularity gives the
boundary conditions
\begin{eqnarray}
\psi(r_{+})=\frac{f^\prime(r_{+})}{m^{2}}\psi^\prime(r_{+})\,,\hspace{0.5cm}
\phi(r_{+})=0\,. \label{horizon}
\end{eqnarray}
At the asymptotic AdS boundary $r\rightarrow\infty$, the solutions
behave like
\begin{eqnarray}
\psi=\frac{\psi_{-}}{r^{\lambda_{-}}}+\frac{\psi_{+}}{r^{\lambda_{+}}}\,,\hspace{0.5cm}
\phi=\mu-\frac{\rho}{r^{d-3}}\,, \label{infinity}
\end{eqnarray}
with
\begin{eqnarray}
\lambda_\pm=\frac{1}{2}[(d-1)\pm\sqrt{(d-1)^{2}+4m^{2}L^2}~], \label{lambda}
\end{eqnarray}
where $\mu$ and $\rho$ are interpreted as the chemical potential and charge density in the dual field theory respectively. It should be noted that, provided $\lambda_{-}$ is larger than the unitarity bound, the coefficients $\psi_{-}$ and $\psi_{+}$ both multiply normalizable modes of the scalar field equations and they correspond to the vacuum expectation values $\langle{\mathcal{O}_{-}}\rangle=\psi_{-}$, $\langle{\mathcal{O}_{+}}\rangle=\psi_{+}$ of an operator $\mathcal{O}$ dual to the scalar field according to the AdS/CFT correspondence. Just as in Refs. \cite{HartnollPRL101,HartnollJHEP12}, we can impose boundary condition that either $\psi_{+}$ or $\psi_{-}$ vanishes. For simplicity, we will scale $L=1$ in the following calculation.

\subsection{The condensation of the scalar operators}

Using the shooting method, we can solve the equations of motion (\ref{BHPsi}) and (\ref{BHPhi}) numerically and then discuss the effects of the nonlinear electrodynamics on the condensation of the scalar operators. For concreteness, we will set $d=4$ and $m^2L^2=-2$. It should be noted that the other choices of the mass of the scalar field and the dimensionality of the spacetime will not qualitatively modify our results.

\begin{figure}[ht]
\includegraphics[scale=0.51]{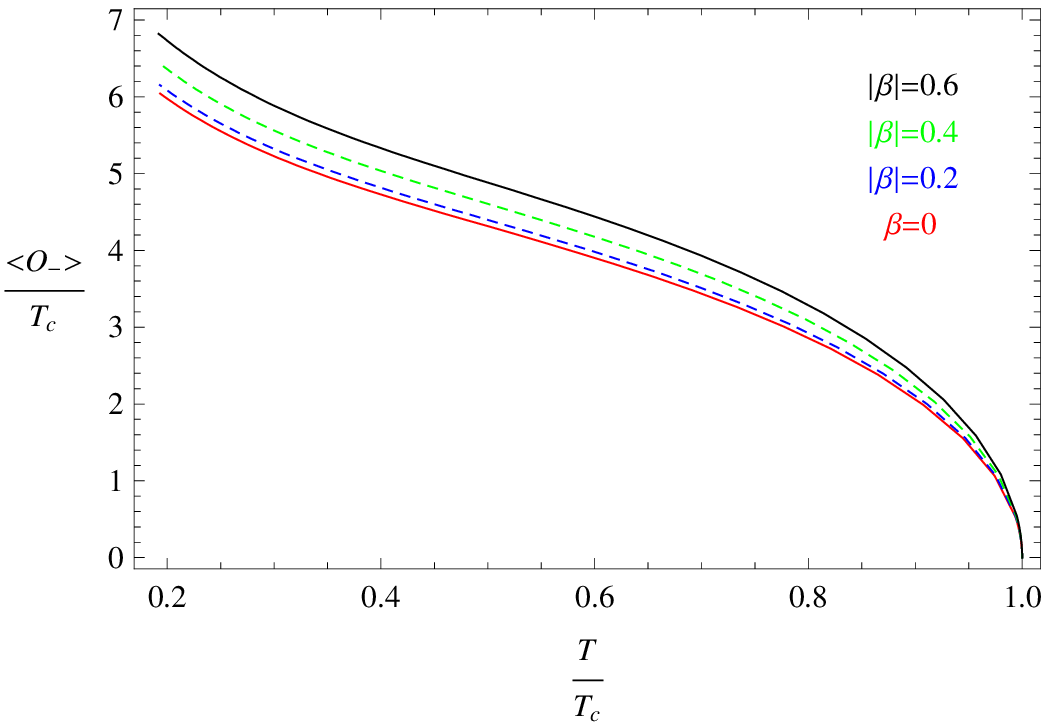}\hspace{0.2cm}%
\includegraphics[scale=0.51]{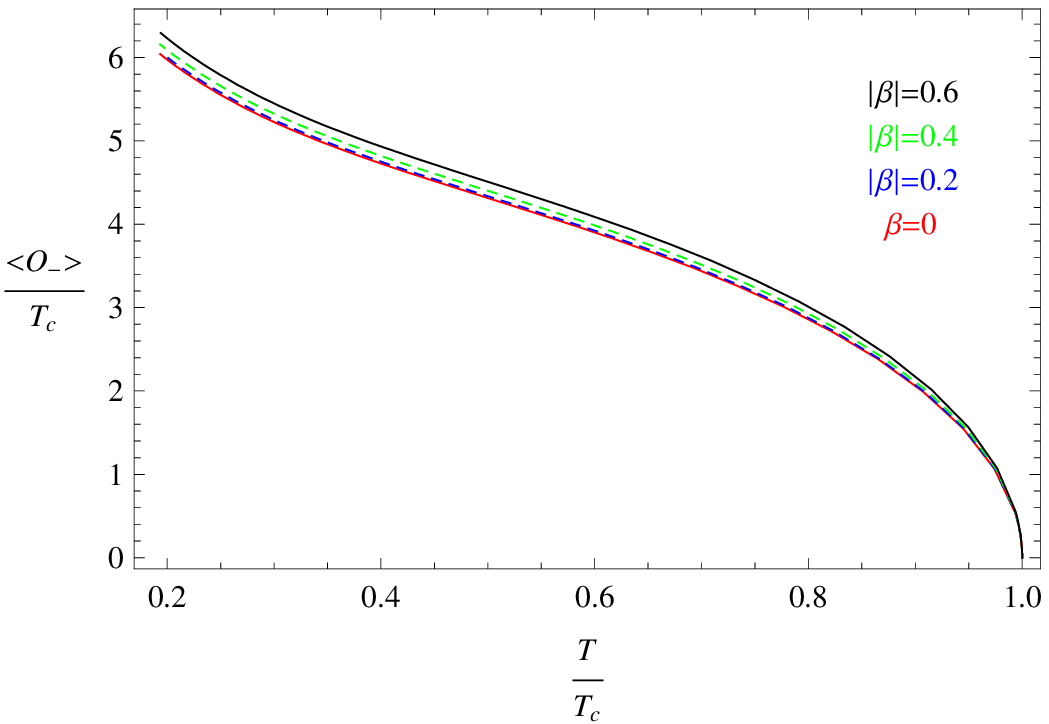}\hspace{0.2cm}%
\includegraphics[scale=0.51]{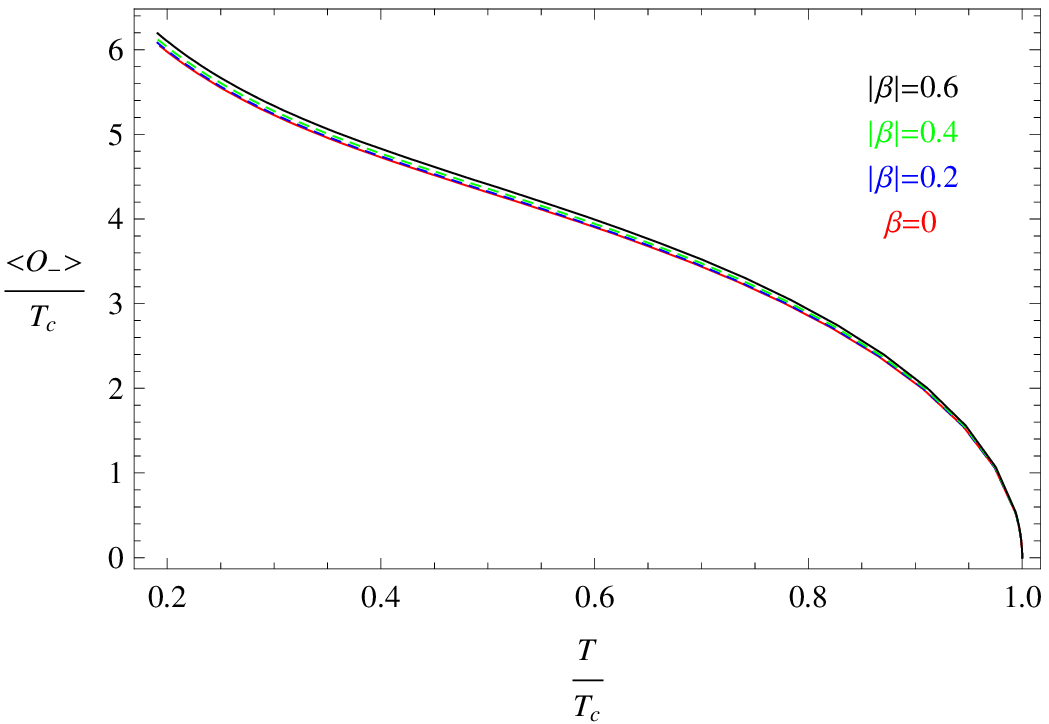}\\ \vspace{0.0cm}
\includegraphics[scale=0.51]{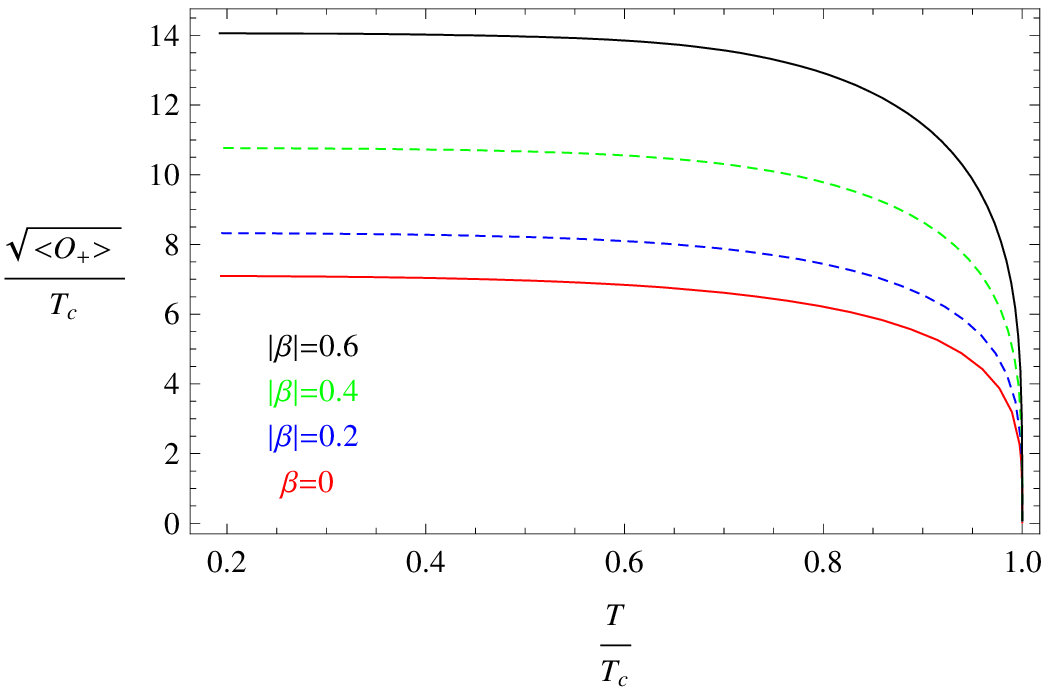}\hspace{0.2cm}%
\includegraphics[scale=0.51]{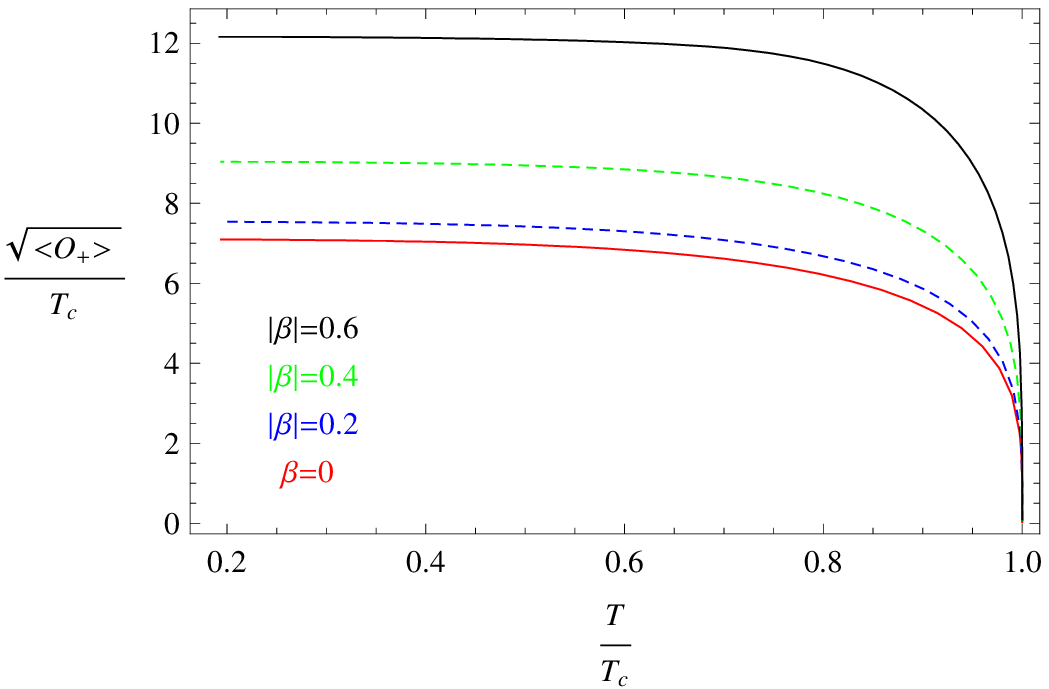}\hspace{0.2cm}%
\includegraphics[scale=0.51]{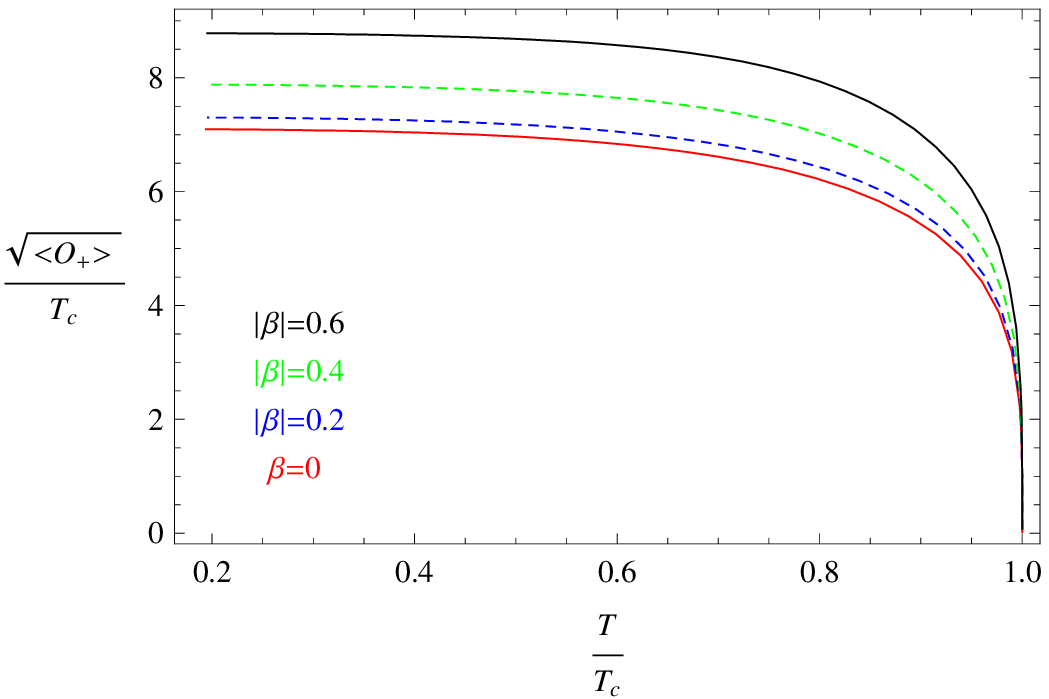}\\ \vspace{0.0cm}
\caption{\label{CondNLE} (color online) The condensates of the scalar operators $\mathcal{O}_{-}$ and $\mathcal{O}_{+}$ with the ENE (left two panels), BINE (middle two panels) and LNE (right two panels) as a function of temperature for the mass of the scalar field $m^2L^2=-2$ in $d=4$ dimension. In each panel, the four lines from bottom to top correspond to increasing absolute value of the nonlinearity parameter, i.e., $|\beta|=0$ (red), $0.2$ (blue), $0.4$ (green) and $0.6$ (black) respectively. }
\end{figure}

Changing the nonlinearity parameter $\beta$, we present in Fig. \ref{CondNLE} the condensates of the scalar operators $\mathcal{O}_{-}$ and $\mathcal{O}_{+}$ with the ENE (left two panels), BINE (middle two panels) and LNE (right two panels) as a function of temperature for the mass of the scalar field $m^2L^2=-2$ in $d=4$ dimension. It is found that the curves for the operator $\mathcal{O}_{+}$ in the bottom three panels of the Fig. \ref{CondNLE} have similar behavior to the BCS theory for different $\beta$, where the condensate goes to a constant at zero temperature. However, the curves for the operator $\mathcal{O}_{-}$ in the top three panels will diverge at low temperature, which are similar to that for the usual Maxwell electrodynamics in the probe limit neglecting backreaction of the spacetime \cite{HartnollPRL101}. Obviously, similar to the cases of BINE and LNE, the holographic superconductors still exist even we consider the exponential form of nonlinear electrodynamics.

From the left two panels of Fig. \ref{CondNLE}, we observe that the increasing absolute value of the nonlinearity parameter $|\beta|$ makes the condensation gap larger for both scalar operators $\mathcal{O}_{-}$ and $\mathcal{O}_{+}$ with the ENE, which is similar to the cases of BINE and LNE. This means that the scalar hair is harder to be formed in the nonlinear electrodynamics, which agrees well with the results given in \cite{JS2010,JPCPLB}. In fact, the tables \ref{TcD4Zheng} and \ref{TcD4Fu} show that the critical temperature $T_{c}$ for the operators $\mathcal{O}_{-}$ and $\mathcal{O}_{+}$ decreases as the absolute value of the nonlinearity parameter $|\beta|$ increases, which agrees well with the finding in Fig. \ref{CondNLE}. This behavior is reminiscent of that seen for the Gauss-Bonnet holographic superconductors, where the higher curvature corrections make condensation harder to form \cite{Gregory,Pan-Wang,Ge-Wang}, so we conclude that the ENE, BINE and LNE corrections to the usual Maxwell field and the curvature corrections share some similar features for the condensation of the scalar operators.

\begin{table}[ht]
\caption{\label{TcD4Zheng} The critical temperature $T_{c}$ for the operator $\mathcal{O}_{-}$ with different values of $\beta$ for $d=4$ and $m^2L^2=-2$. We have set $\rho=1$ in the table.}
\begin{tabular}{c c c c c c c c}
         \hline
$|\beta|$ & 0 & 0.1 & 0.2 & 0.3 & 0.4 & 0.5 & 0.6
        \\
        \hline
~~~~ENE~~~~&~~~~~$0.2255$~~~~~&~~~~~$0.2247$~~~~~&~~~~~$0.2225$~~~~~&
~~~~~$0.2217$~~~~&~~~~~$0.2152$~~~~&~~~~~$0.2107$~~~~&~~~~~$0.2061$~~~~
          \\
~~~~BINE~~~~&~~~~~$0.2255$~~~~~&~~~~~$0.2253$~~~~~&~~~~~$0.2247$~~~~~&
~~~~~$0.2237$~~~~&~~~~~$0.2223$~~~~&~~~~~$0.2206$~~~~&~~~~~$0.2185$~~~~
          \\
~~~~LNE~~~~&~~~~~$0.2255$~~~~~&~~~~~$0.2254$~~~~~&~~~~~$0.2251$~~~~~&
~~~~~$0.2246$~~~~&~~~~~$0.2239$~~~~&~~~~~$0.2231$~~~~&~~~~~$0.2220$~~~~
          \\
        \hline
\end{tabular}
\end{table}

\begin{table}[ht]
\caption{\label{TcD4Fu} The critical temperature $T_{c}$ for the operator $\mathcal{O}_{+}$ with different values of $\beta$ for $d=4$ and $m^2L^2=-2$. We have set $\rho=1$ in the table.}
\begin{tabular}{c c c c c c c c}
         \hline
$|\beta|$ & 0 & 0.1 & 0.2 & 0.3 & 0.4 & 0.5 & 0.6
        \\
        \hline
~~~~ENE~~~~&~~~~~$0.1184$~~~~~&~~~~~$0.1120$~~~~~&~~~~~$0.0994$~~~~~&
~~~~~$0.0861$~~~~&~~~~~$0.0737$~~~~&~~~~~$0.0627$~~~~&~~~~~$0.0531$~~~~
          \\
~~~~BINE~~~~&~~~~~$0.1184$~~~~~&~~~~~$0.1165$~~~~~&~~~~~$0.1110$~~~~~&
~~~~~$0.1024$~~~~&~~~~~$0.0914$~~~~&~~~~~$0.0790$~~~~&~~~~~$0.0662$~~~~
          \\
~~~~LNE~~~~&~~~~~$0.1184$~~~~~&~~~~~$0.1175$~~~~~&~~~~~$0.1148$~~~~~&
~~~~~$0.1108$~~~~&~~~~~$0.1058$~~~~&~~~~~$0.1001$~~~~&~~~~~$0.0939$~~~~
          \\
        \hline
\end{tabular}
\end{table}

On the other hand, comparing with the curves for the scalar operators $\mathcal{O}_{-}$ and $\mathcal{O}_{+}$ in the three types of the nonlinear electrodynamics considered here, we find that the condensation gap of the ENE is larger than that of BINE and LNE for the fixed value of $\beta$ (except the case of $\beta=0$, i.e., the usual Maxwell electrodynamics), which means that the scalar hair is more difficult to be developed in the Exponential form of nonlinear electrodynamics. This is also in good agreement with the results shown in the tables \ref{TcD4Zheng} and \ref{TcD4Fu}, where the critical temperature $T_{c}$ for the operators $\mathcal{O}_{-}$ and $\mathcal{O}_{+}$ with the ENE is smaller than that of BINE and LNE for the fixed value of $\beta$.

\subsection{Electrical conductivity}

Now we want to know the influence of the nonlinear electrodynamics on the electrical conductivity. Assuming that the perturbed Maxwell field has a form $\delta A_{x}=A_{x}(r)e^{-i\omega t}dx$, we obtain the equation of motion for $A_{x}$ in the three types of the nonlinear electrodynamics, which can be used to calculate the conductivity
\begin{eqnarray}
&&A_{x}^{\prime\prime}+\left(\frac{d-4}{r}+\frac{f^\prime}{f}+
4\beta^{2}\phi^{\prime}\phi^{\prime\prime}\right)A_{x}^\prime
+\left(\frac{\omega^2}{f^2}-\frac{2\psi^{2}e^{-2\beta^{2}\phi^{\prime 2}}}{f}\right)A_{x}=0~, \quad {\rm ENE} \nonumber\\
&&A_{x}^{\prime\prime}+\left(\frac{d-4}{r}+\frac{f^\prime}{f}+
\frac{\beta^{2}\phi^{\prime}\phi^{\prime\prime}}{1-\beta^{2}\phi^{\prime 2}}\right)A_{x}^\prime
+\left[\frac{\omega^2}{f^2}-\frac{2\psi^{2}(1-\beta^{2}\phi^{\prime 2})^{1/2}}{f}\right]A_{x}=0~, \quad {\rm BINE} \nonumber\\
&&A_{x}^{\prime\prime}+\left(\frac{d-4}{r}+\frac{f^\prime}{f}+
\frac{2\beta^{2}\phi^{\prime}\phi^{\prime\prime}}{4-\beta^{2}\phi^{\prime 2}}\right)A_{x}^\prime
+\left[\frac{\omega^2}{f^2}-\frac{2\psi^{2}}{f}\left(1-\frac{1}{4}\beta^{2}\phi^{\prime 2}\right)\right]A_{x}=0~, \quad {\rm LNE}
\label{ConductivityEquation}
\end{eqnarray}
For simplicity, we will restrict our study to $d=4$ dimension. Though the above equations are more complicated than that in the usual Maxwell electrodynamics, for all cases considered here the ingoing wave boundary condition near the horizon is still given by
\begin{eqnarray}
A_{x}(r)\sim f(r)^{-\frac{i\omega}{3r_+}},
\end{eqnarray}
and in the asymptotic AdS region ($r\rightarrow\infty$) the general behavior can be written as
\begin{eqnarray}
A_{x}=A^{(0)}+\frac{A^{(1)}}{r}.
\end{eqnarray}
Using the AdS/CFT dictionary, we find that the conductivity of the dual superconductor can be expressed
as \cite{HartnollPRL101,HartnollJHEP12}
\begin{eqnarray}\label{Conductivity}
\sigma=-\frac{iA^{(1)}}{\omega A^{(0)}}\ .
\end{eqnarray}
For different values of the nonlinearity parameter $\beta$, one can obtain the conductivity by solving Eq. (\ref{ConductivityEquation}) numerically. In our following discussion for the three types of the nonlinear electrodynamics, we still focus on the case for the fixed scalar mass $m^2L^2=-2$.

\begin{figure}[H]
\includegraphics[scale=0.5]{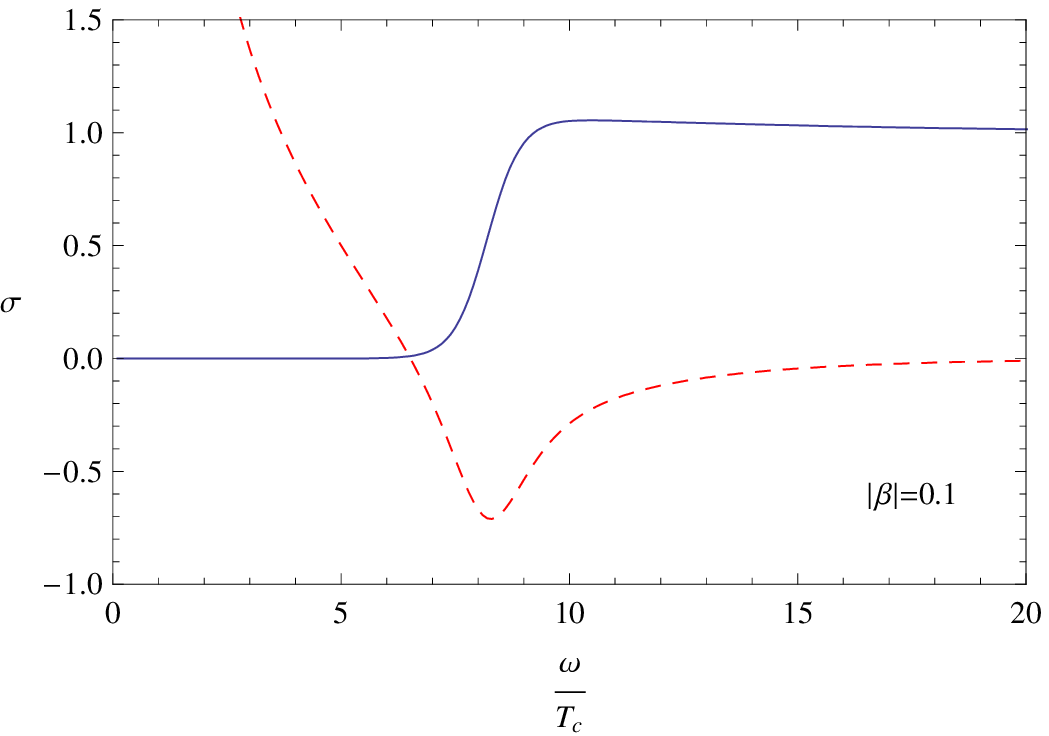}\hspace{0.2cm}%
\includegraphics[scale=0.5]{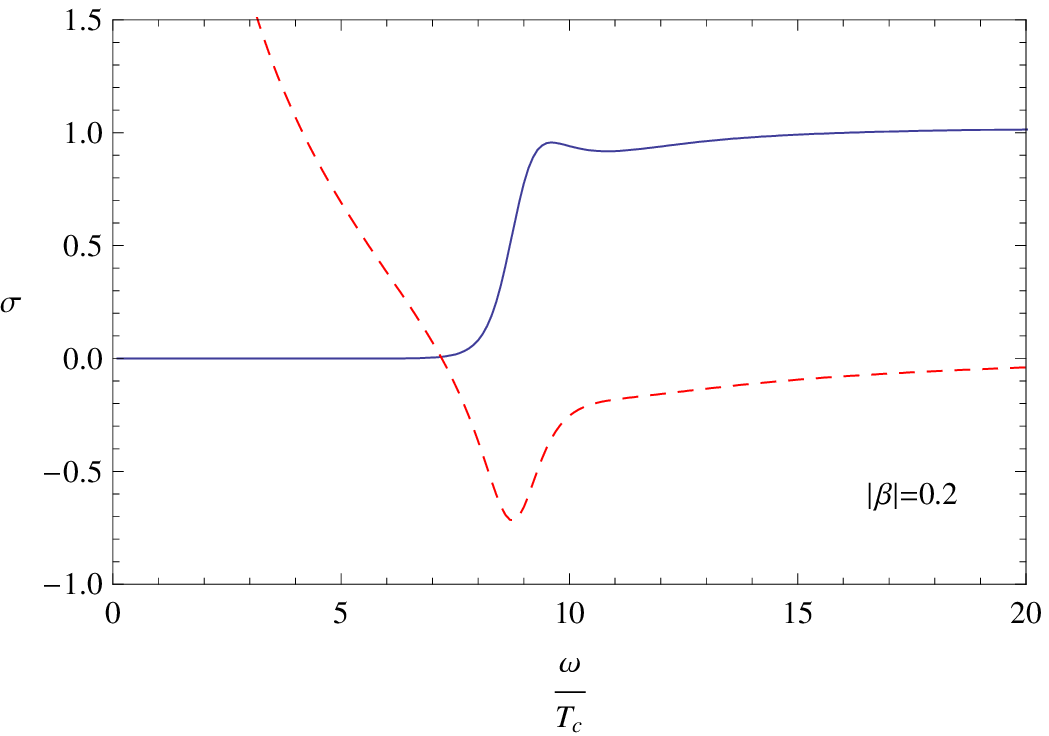}\hspace{0.2cm}%
\includegraphics[scale=0.5]{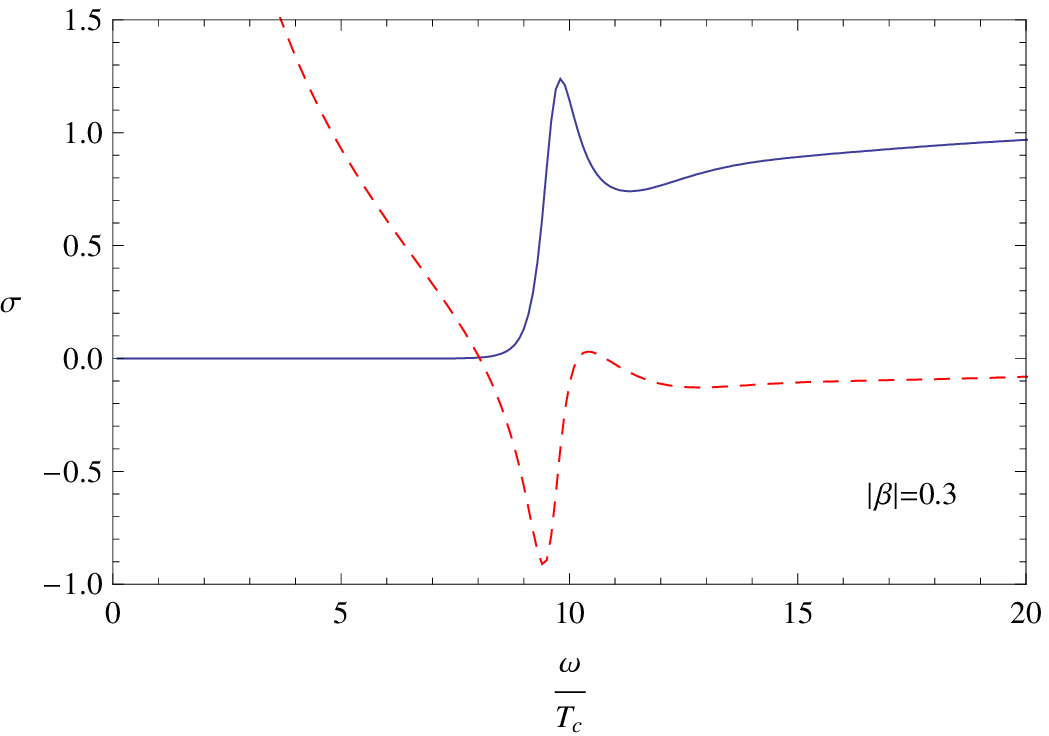}\\ \vspace{0.0cm}
\includegraphics[scale=0.5]{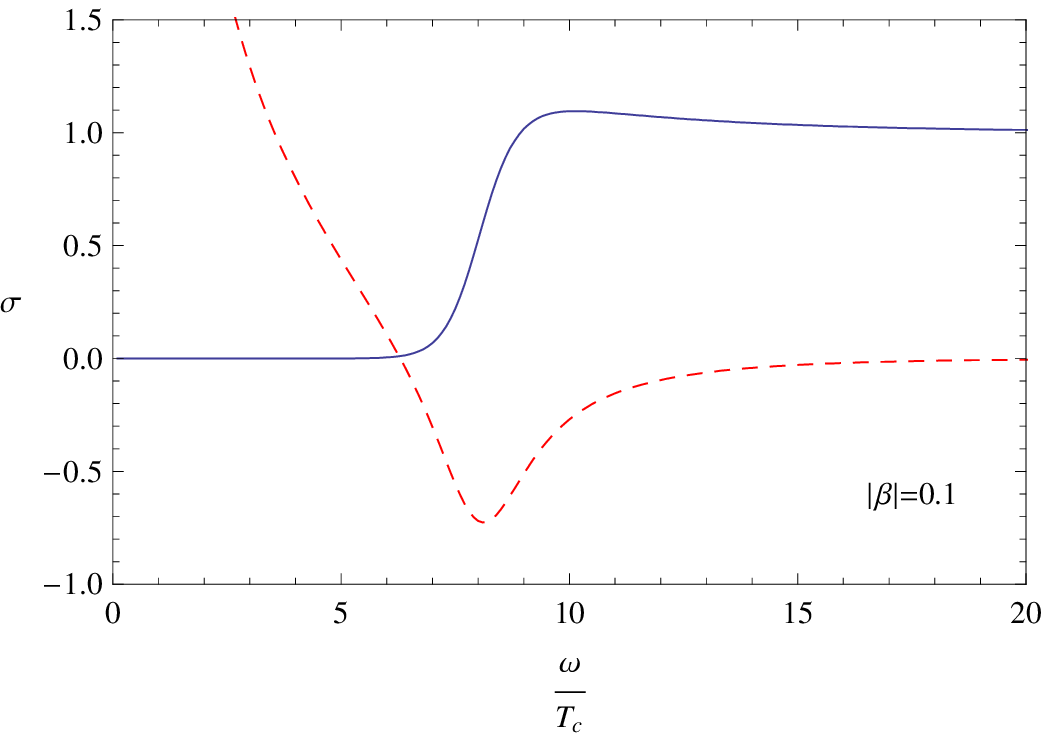}\hspace{0.2cm}%
\includegraphics[scale=0.5]{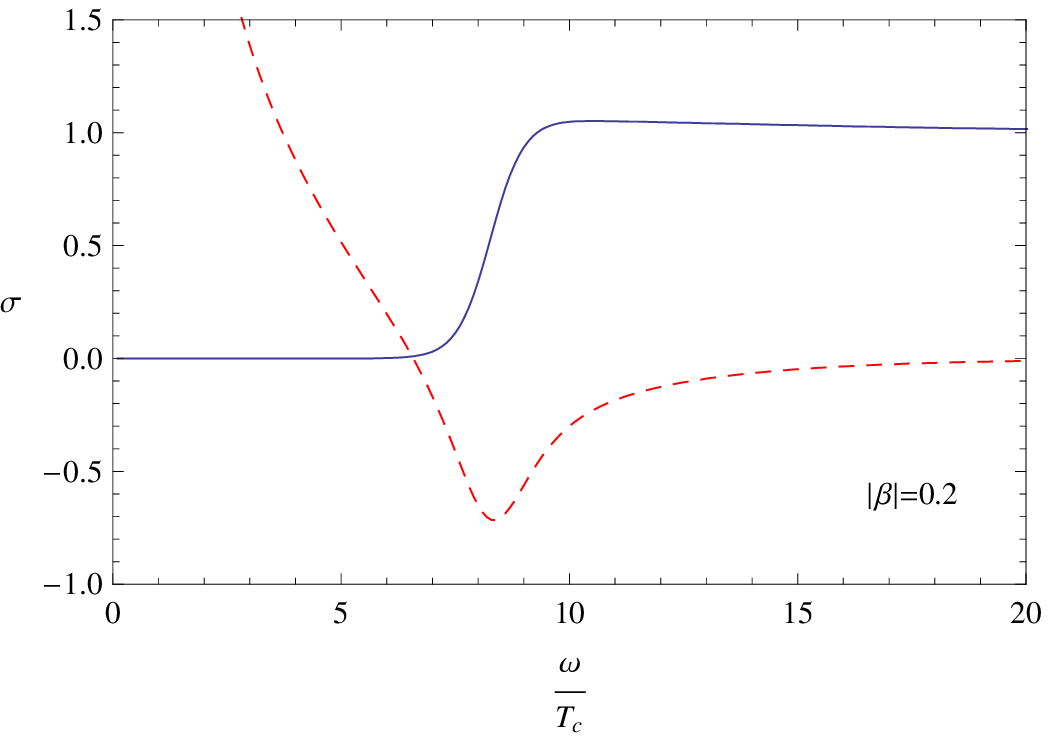}\hspace{0.2cm}%
\includegraphics[scale=0.5]{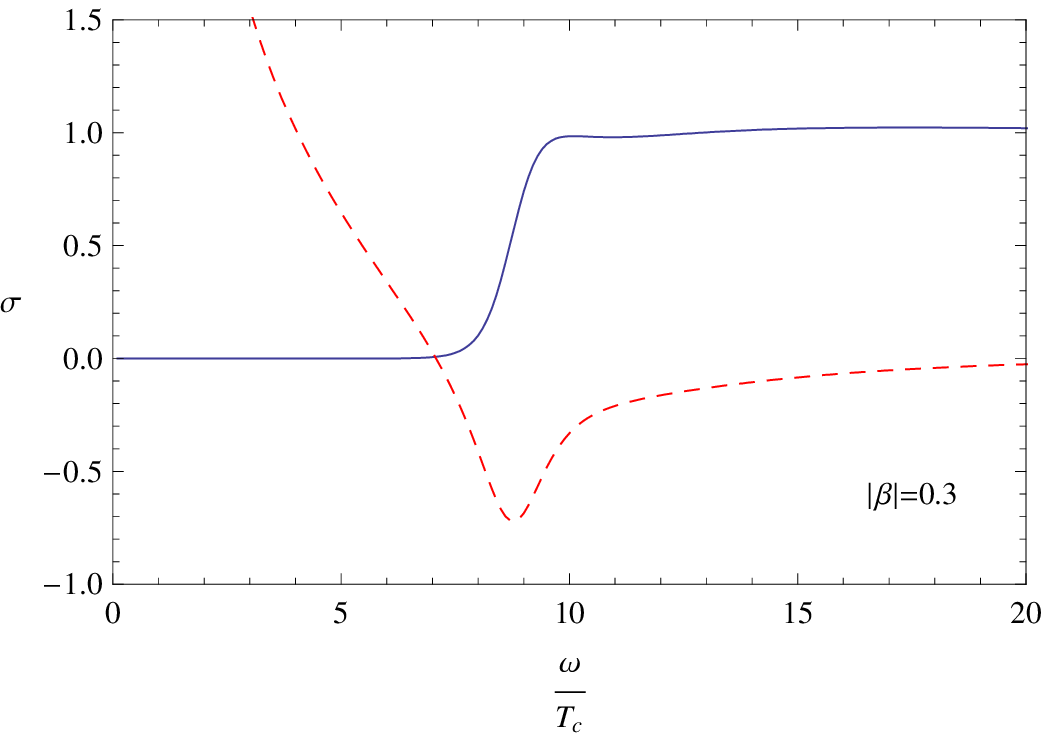}\\ \vspace{0.0cm}
\includegraphics[scale=0.5]{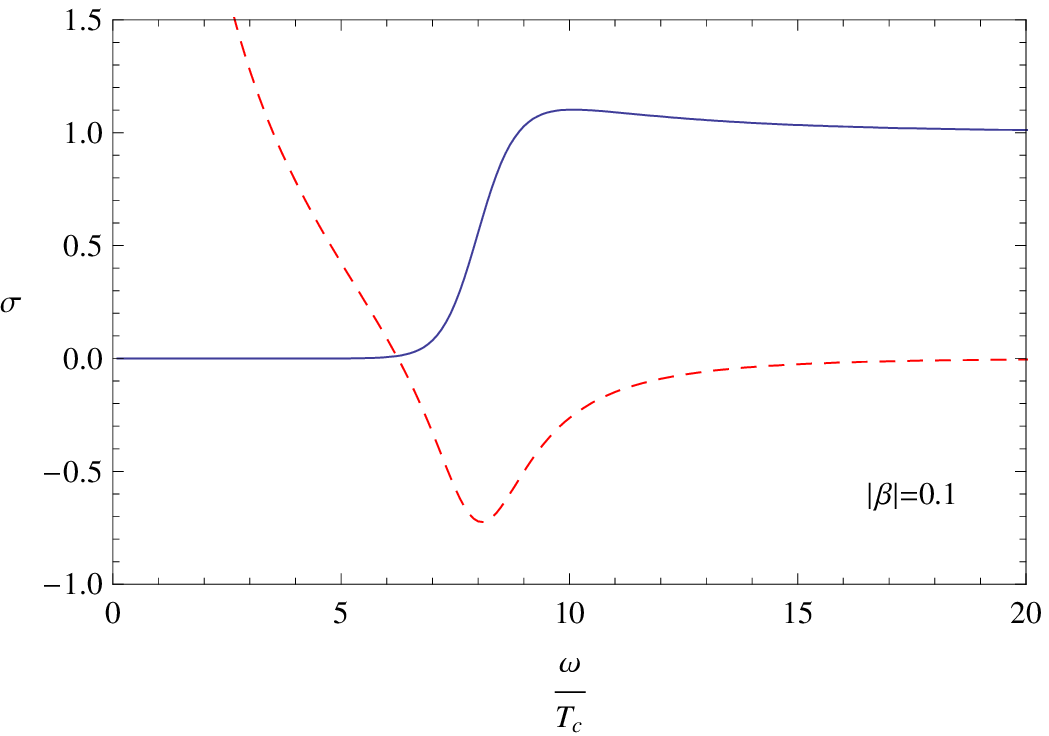}\hspace{0.2cm}%
\includegraphics[scale=0.5]{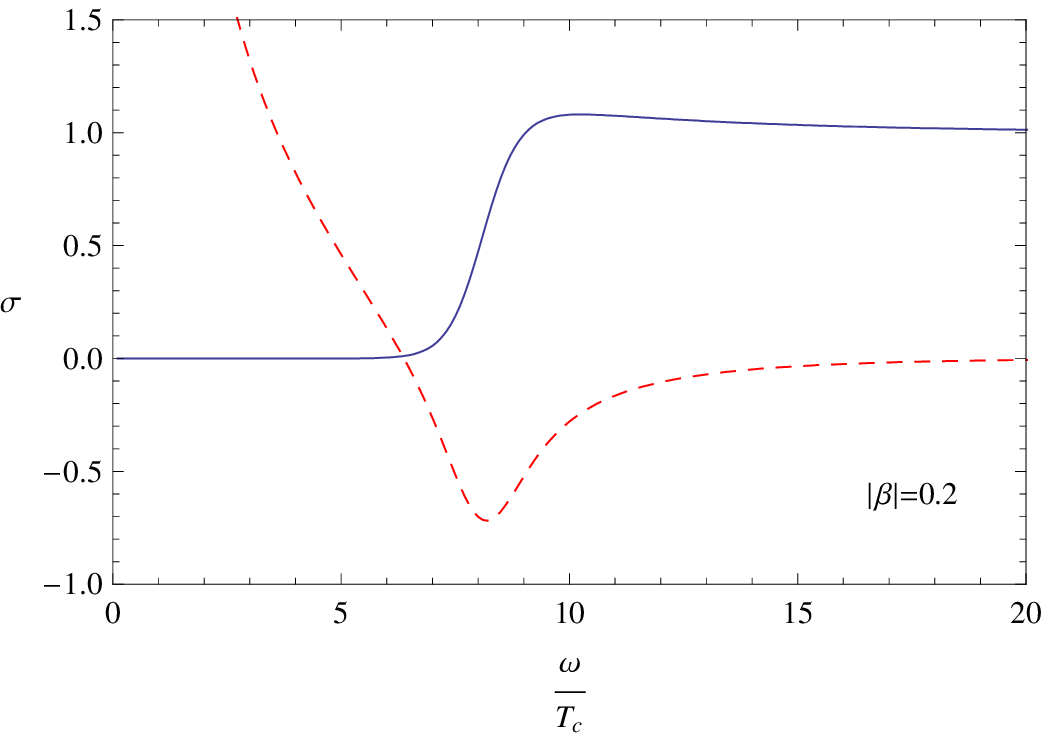}\hspace{0.2cm}%
\includegraphics[scale=0.5]{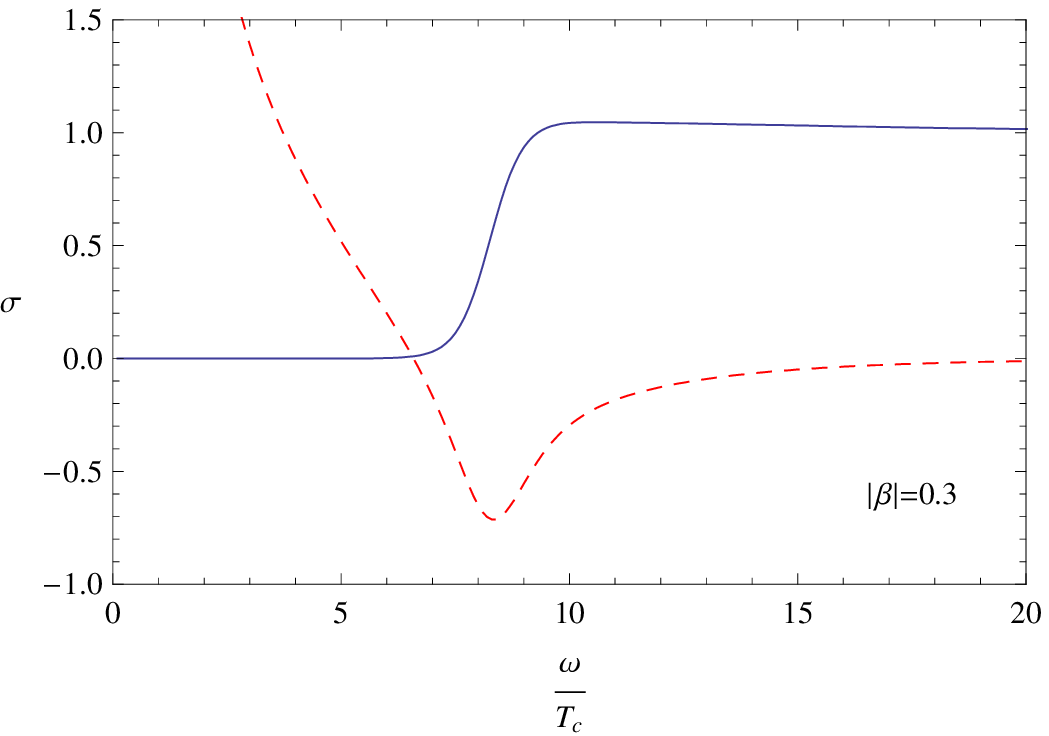}\\ \vspace{0.0cm}
\caption{\label{NLEConductivity} (color online) Conductivity of ($2+1$)-dimensional superconductors with the ENE (top three panels), BINE (middle three panels) and LNE (bottom three panels) for the fixed mass of the scalar field $m^{2}L^{2}=-2$ and different values of the nonlinearity parameter $\beta$.}
\end{figure}

For the three types of the nonlinear electrodynamics, we plot the frequency dependent conductivity obtained by solving Eq. (\ref{ConductivityEquation}) numerically for $|\beta|=0.1$, $0.2$ and $0.3$ at temperature $T/T_{c}\approx0.2$ in Fig. \ref{NLEConductivity}. In each panel, the blue (solid) line and red (dashed) line represent the real part and imaginary part of the conductivity $\sigma(\omega)$ respectively. For all cases considered here, we find a gap in the conductivity with the gap frequency $\omega_{g}$ which becomes larger when we increase the values of $|\beta|$. Also, for increasing values of $|\beta|$, we have larger deviations from the value $\omega_g/T_c\approx 8$, which has been presented in table \ref{NLEConductivityTc}. This shows that, similar to the effect of the Gauss-Bonnet coupling, the nonlinear electrodynamics really change the ratio in the gap frequency $\omega_g/T_c\approx 8$ which was claimed to be universal in \cite{HorowitzPRD78}.

\begin{table}[ht]
\caption{\label{NLEConductivityTc} The ratio $\omega_{g}/T_{c}$ with fixed values of $\beta$ for $d=4$ and $m^2L^2=-2$. In order to compare with the result of the usual Maxwell electrodynamics, we also present the ratio $\omega_{g}/T_{c}$ of $\beta=0$.}
\begin{tabular}{ccccc}
         \hline
$|\beta|$ &~~~~~0~~~~~&~~~~~0.1~~~~~&~~~~~0.2~~~~~&~~~~~0.3~~~~~
          \\
        \hline
~~~~ENE~~~~&~~~~~$8.1$~~~~~&~~~~~$8.3$~~~~~&~~~~~$8.8$~~~~~&~~~~~$9.4$~~~~~
          \\
~~~~BINE~~~~&~~~~~$8.1$~~~~~&~~~~~$8.2$~~~~~&~~~~~$8.4$~~~~~&~~~~~$8.8$~~~~~
          \\
~~~~LNE~~~~&~~~~~$8.1$~~~~~&~~~~~$8.1$~~~~~&~~~~~$8.2$~~~~~&~~~~~$8.4$~~~~~
          \\
         \hline
\end{tabular}
\end{table}

On the other hand, from Fig. \ref{NLEConductivity} and table \ref{NLEConductivityTc} we can see that the deviation from the value $\omega_g/T_c\approx 8$ of the ENE is larger than that of BINE and LNE for the fixed value of $\beta$ (except the case of $\beta=0$, i.e., the usual Maxwell electrodynamics), which is consistent with the previous findings in the behaviors of the condensation gap for the scalar operators $\mathcal{O}_{-}$ and $\mathcal{O}_{+}$. Thus, we can conclude that the ENE has stronger effects on the condensation formation and conductivity for the holographic superconductors than the BINE and LNE.

\section{Holographic superconductor/insulator transitions with the nonlinear electrodynamics}

Since the properties of holographic dual models with the nonlinear electrodynamics in the background of AdS black hole strongly depend on the nonlinearity parameter, i.e., the higher nonlinear electrodynamics corrections will make the condensation harder to form, it seems to be an interesting study to consider the influences of the nonlinear electrodynamics on the holographic dual to the AdS soliton.

\subsection{Phase transition between the superconductor and insulator in the AdS soliton}

Now we are in a position to study the superconducting phase for a nonlinear electrodynamic field coupled with a charged scalar field in the $d$-dimensional AdS soliton. Making use of two wick rotations for the AdS Schwarzschild black hole given in (\ref{BH metric}), we can get the metric of the AdS soliton
\begin{eqnarray}\label{soliton}
ds^2=-r^{2}dt^{2}+\frac{dr^2}{f(r)}+f(r)d\varphi^2+r^{2}dx_{j}dx^{j},
\end{eqnarray}
with
\begin{eqnarray}
f(r)=\frac{r^2}{L^2}\left(1-\frac{r_{s}^{d-1}}{r^{d-1}}\right).
\end{eqnarray}
This spacetime does not have any horizon but a conical singularity at $r=r_{s}$. For the smoothness at the tip $r_{s}$, we can impose a period $\beta_{s}=\frac{4\pi L^{2}}{(d-1)r_{s}}$ for the coordinate $\varphi$ to remove the singularity.

Using the generalized action (\ref{System}) for the nonlinear electrodynamic field coupled with the charged scalar field, we can obtain the equation of motion for the scalar field $\psi$ in the probe limit
\begin{eqnarray}
\psi^{\prime\prime}+\left(\frac{d-2}{r}+\frac{f^\prime}{f}\right)\psi^\prime
+\left(\frac{\phi^2}{r^2f}-\frac{m^2}{f}\right)\psi=0\,,
\label{solitonPsi}
\end{eqnarray}
and the following equations of motion for the gauge field $\phi$
\begin{eqnarray}
&&\left(1+\frac{4\beta^{2}f}{r^{2}}\phi^{\prime 2}\right)\phi^{\prime\prime}+\left[\frac{d-4}{r}+\frac{f^\prime}{f}
+\frac{2\beta^{2}f}{r^{2}}\left(\frac{f^\prime}{f}-\frac{2}{r}\right)\phi^{\prime 2}\right]\phi^\prime-\frac{2\psi^{2}}{f}e^{-2\beta^{2}f\phi^{\prime 2}/r^{2}}\phi=0~, \quad {\rm ENE} \nonumber\\
&&\phi^{\prime\prime}+\left[\frac{d-4}{r}+\frac{f^\prime}{f}
-\frac{\beta^{2}f}{r^{2}}\left(\frac{f^\prime}{2f}+\frac{d-3}{r}\right)\phi^{\prime 2}\right]\phi^\prime-\frac{2\psi^{2}}{f}\left(1-\frac{\beta^{2}f}{r^{2}}\phi^{\prime 2}\right)^{3/2}\phi=0~, \quad {\rm BINE} \nonumber\\
&&\left(1+\frac{\beta^{2}f}{4r^{2}}\phi^{\prime 2}\right)\phi^{\prime\prime}+\left[\frac{d-4}{r}+\frac{f^\prime}{f}
-\frac{(d-2)\beta^{2}f}{4r^{3}}\phi^{\prime 2}\right]\phi^\prime-\frac{2\psi^{2}}{f}\left(1-\frac{\beta^{2}f}{4r^{2}}\phi^{\prime 2}\right)^{2}\phi=0~, \quad {\rm LNE} \label{solitonPhi}
\end{eqnarray}
When $\beta\rightarrow0$, Eqs. (\ref{solitonPsi}) and  (\ref{solitonPhi}) reduce to the standard holographic superconductor/insulator phase transition model investigated in \cite{Nishioka-Ryu-Takayanagi}. Using appropriate boundary conditions in the asymptotic region $r\rightarrow\infty$ and at the tip $r=r_{s}$, we can solve these equations numerically. Near the AdS boundary $r\rightarrow\infty$, the solutions for $\psi$ and $\phi$ have the same form just as Eq. (\ref{infinity}). But at the tip $r=r_{s}$, the solutions satisfy
\begin{eqnarray}
\psi=\tilde{\psi}_{0}+\tilde{\psi}_{1}(r-r_{s})+\tilde{\psi}_{2}(r-r_{s})^{2}+\cdots\,, \nonumber \\
\phi=\tilde{\phi}_{0}+\tilde{\phi}_{1}(r-r_{s})+\tilde{\phi}_{2}(r-r_{s})^{2}+\cdots\,,
\label{SolitonBoundary}
\end{eqnarray}
where $\tilde{\psi}_{i}$ and $\tilde{\phi}_{i}$ ($i=0,1,2,\cdots$) are integration constants. In order to keep every physical quantity finite, we have imposed the Neumann-like boundary conditions \cite{Nishioka-Ryu-Takayanagi}. For concreteness and simplicity, we will take $d=5$ and scale $r_{s}=1$ in the following calculation just as in \cite{Nishioka-Ryu-Takayanagi}.

It is well-known that the normal insulator phase is described by a pure AdS soliton solution. However, the pure AdS soliton background will become unstable to develop a scalar hair when the chemical potential is bigger than a critical value $\mu_{c}$ \cite{Nishioka-Ryu-Takayanagi}. Thus, a new and stable AdS soliton solution with nontrivial scalar field appears, which is dual to a superconducting phase. In this way, one can realize the phase transition between the superconductor and insulator phases around the critical chemical potential $\mu_{c}$. We will examine the effect of the nonlinearity parameter $\beta$ on $\mu_{c}$ numerically.

\begin{figure}[ht]
\includegraphics[scale=0.51]{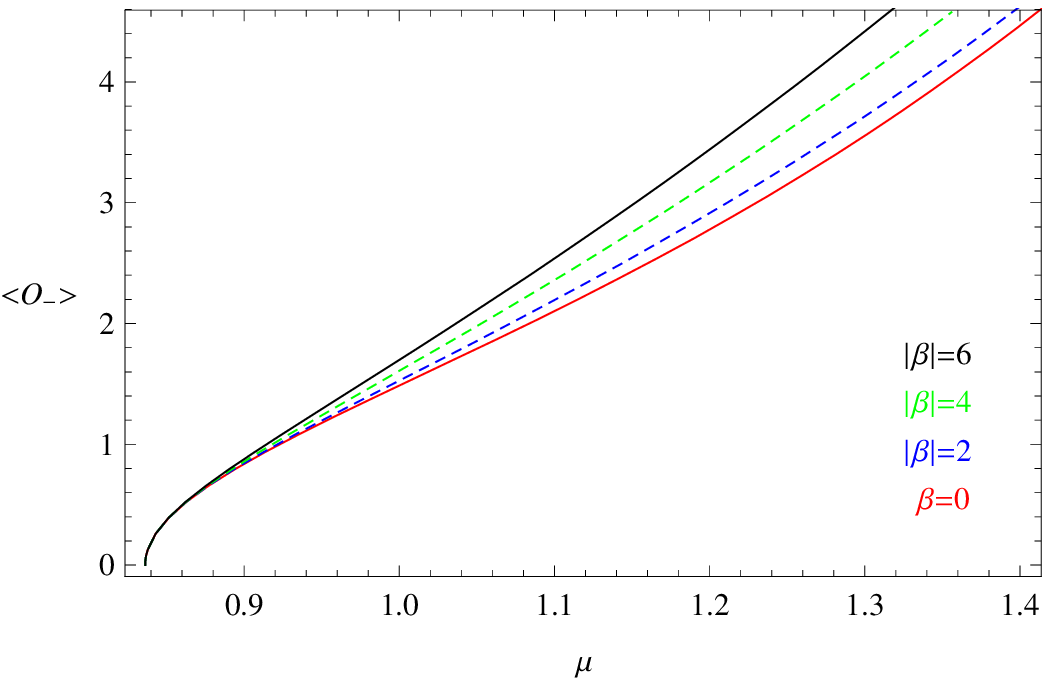}\hspace{0.2cm}%
\includegraphics[scale=0.51]{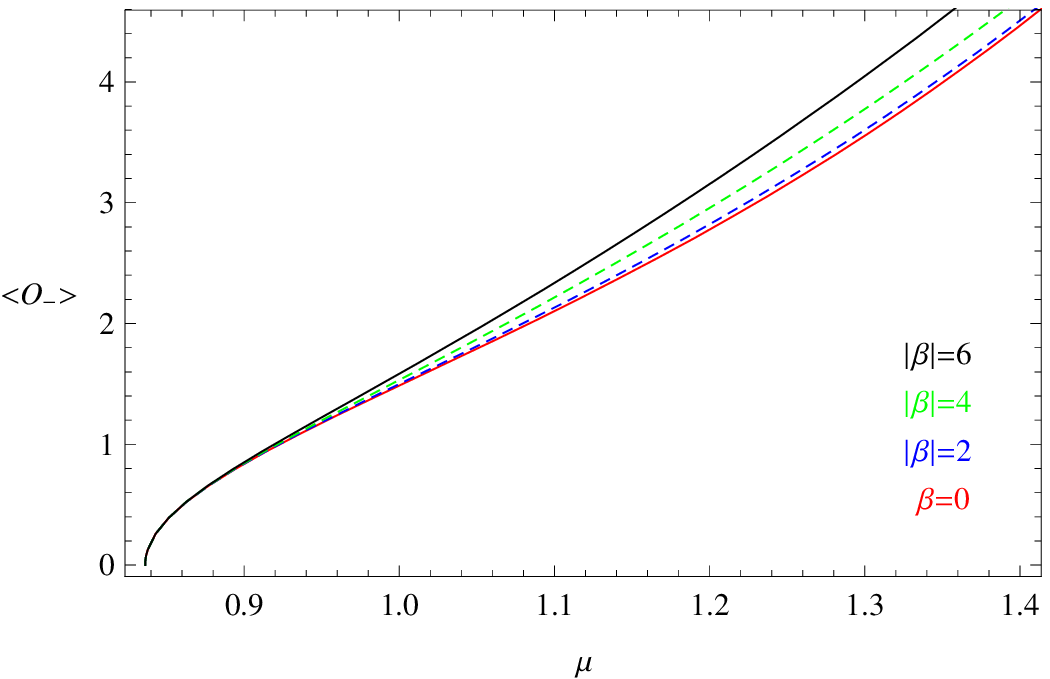}\hspace{0.2cm}%
\includegraphics[scale=0.51]{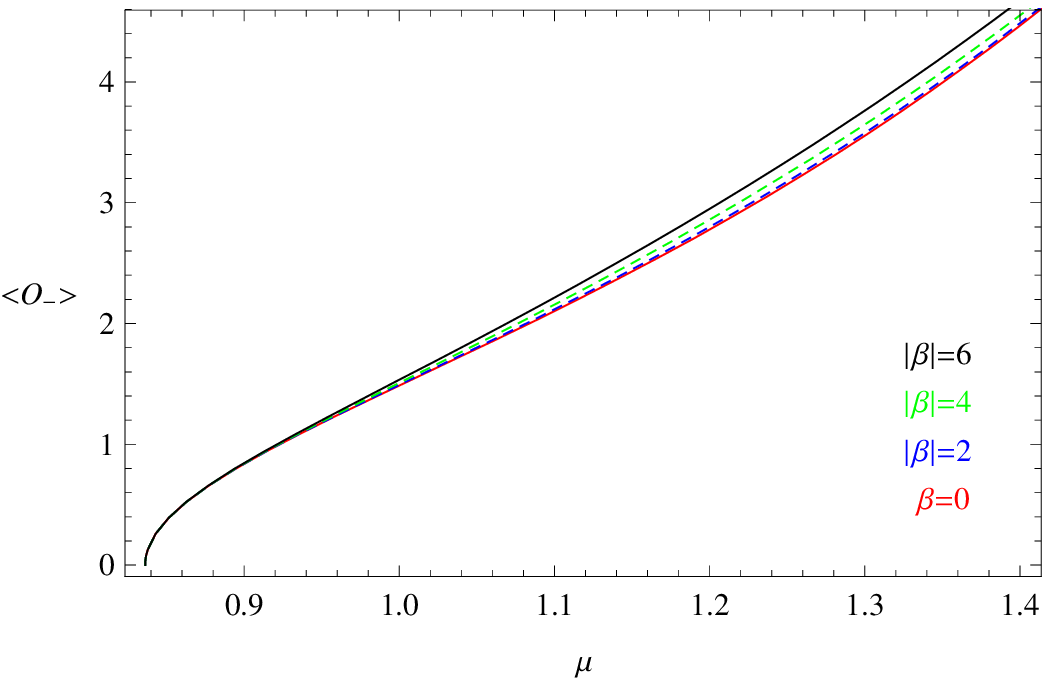}\\ \vspace{0.0cm}
\includegraphics[scale=0.51]{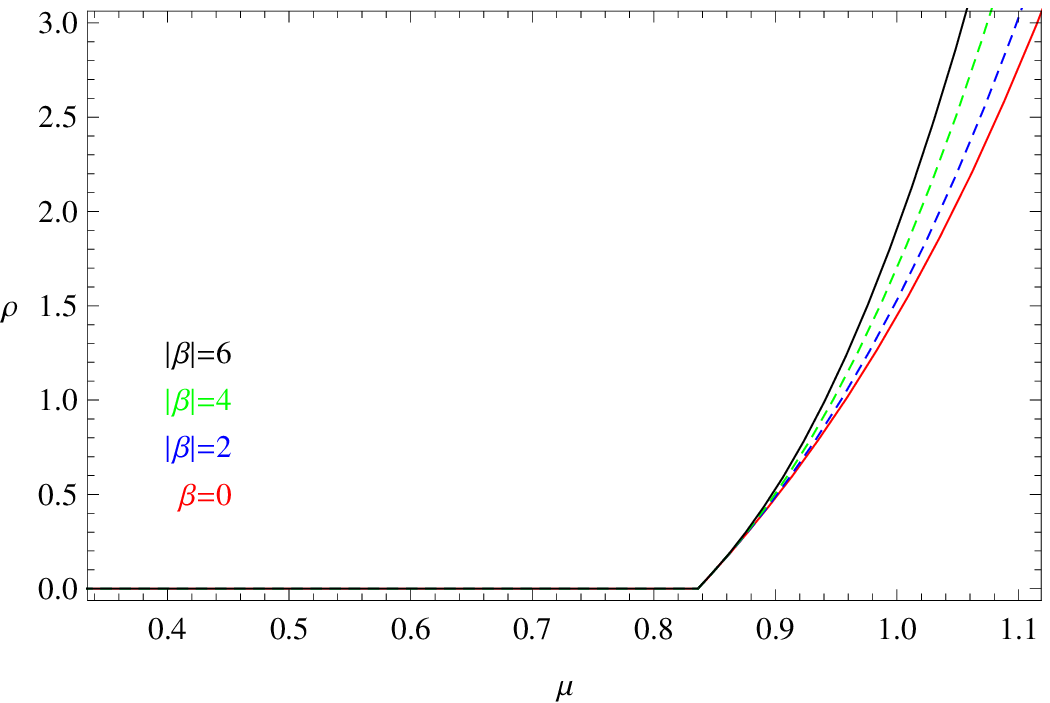}\hspace{0.2cm}%
\includegraphics[scale=0.51]{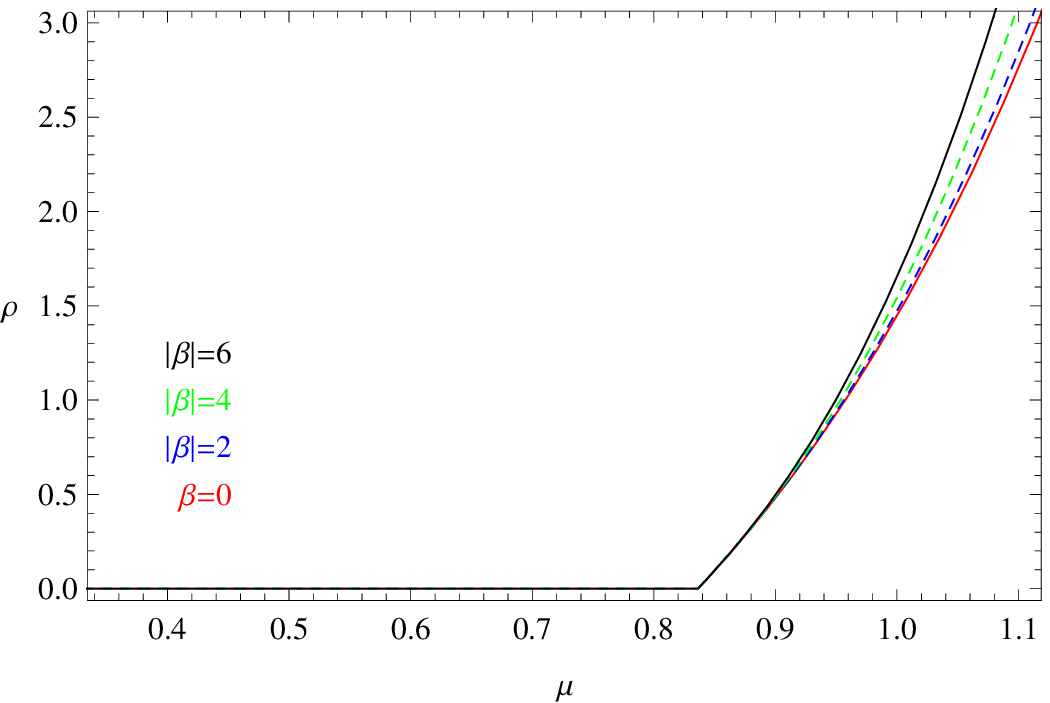}\hspace{0.2cm}%
\includegraphics[scale=0.51]{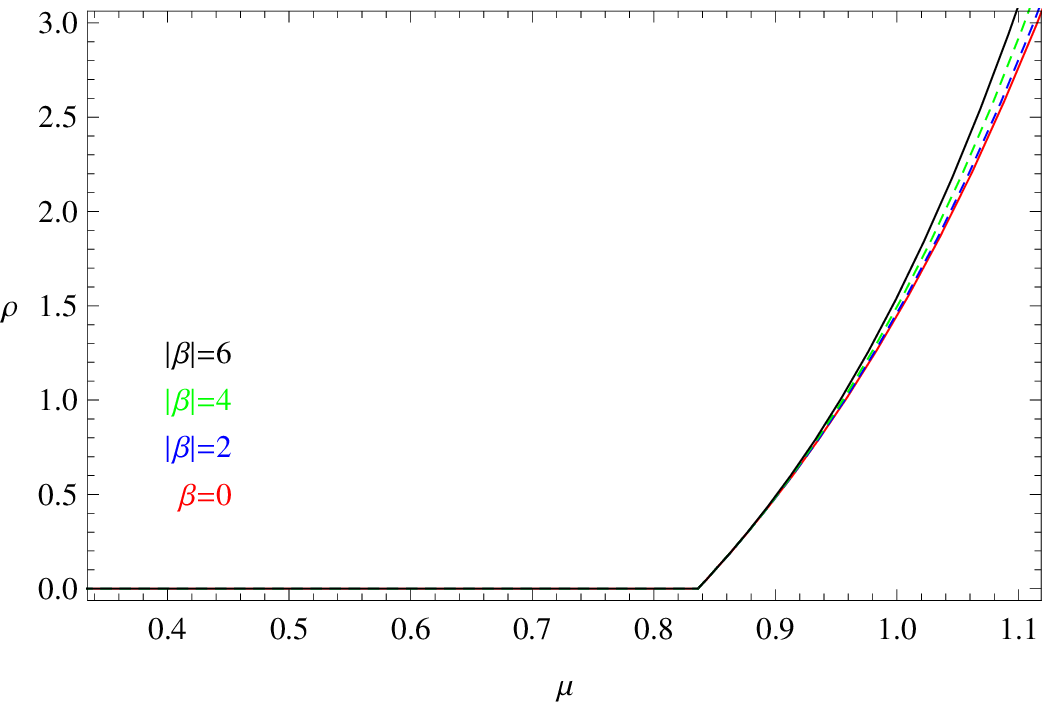}\\ \vspace{0.0cm}
\caption{\label{SolitonNLEF375} (color online) The condensate of the operator $\cal O_{-}$ and charge density $\rho$ with the ENE (left two panels), BINE (middle two panels) and LNE (right two panels) as a function of the chemical potential $\mu$ for different nonlinearity parameters $\beta$ with the mass of the scalar field $m^{2}L^{2}=-15/4$ for the holographic superconductor and insulator model. In each panel, the four lines from bottom to top correspond to increasing absolute value of the nonlinearity parameter, i.e., $|\beta|=0$ (red), $2$ (blue), $4$ (green) and $6$ (black) respectively.}
\end{figure}

\begin{figure}[ht]
\includegraphics[scale=0.51]{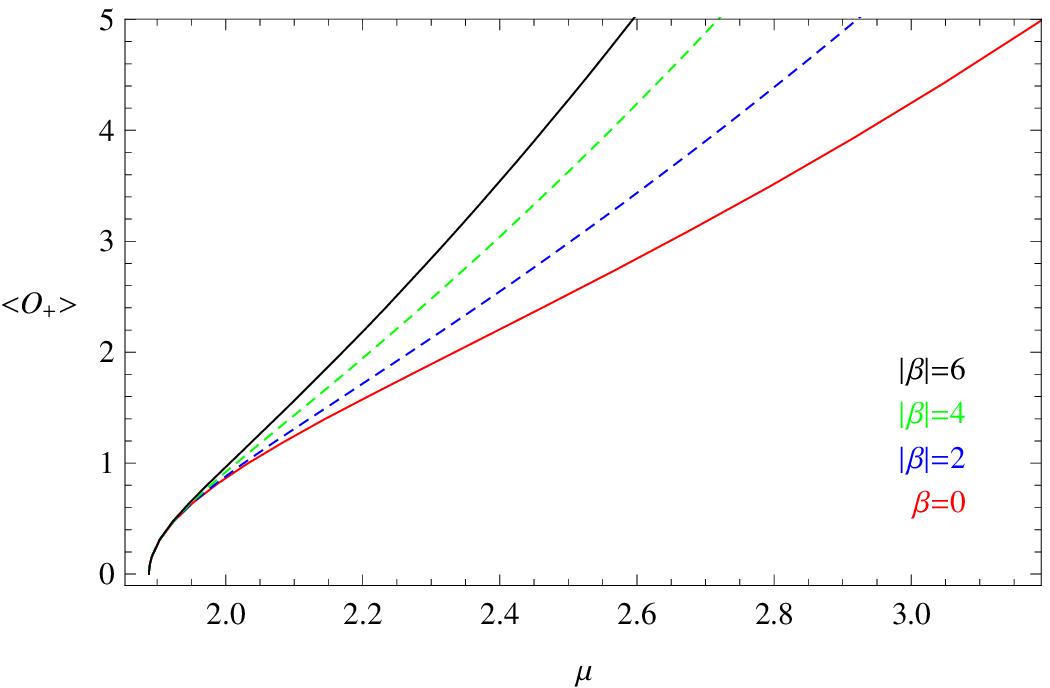}\hspace{0.2cm}%
\includegraphics[scale=0.51]{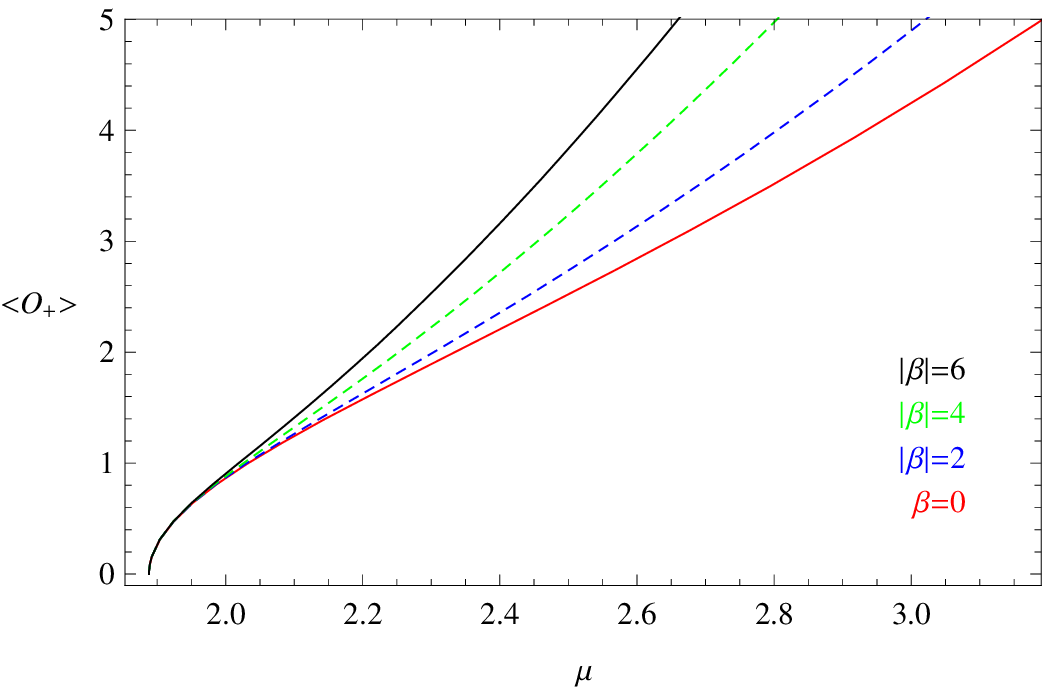}\hspace{0.2cm}%
\includegraphics[scale=0.51]{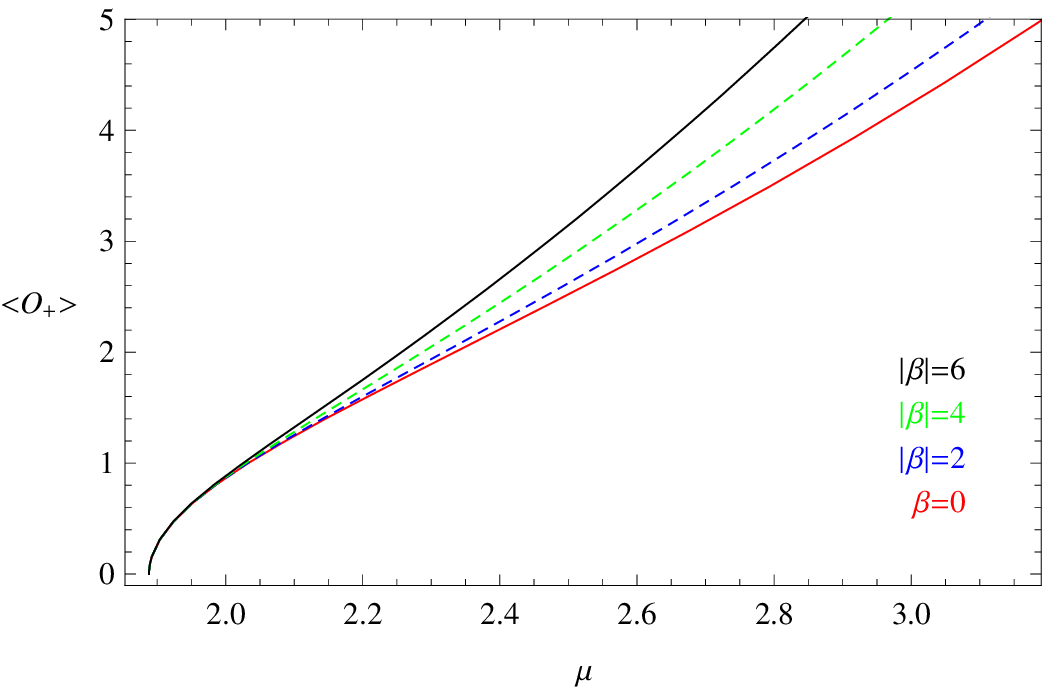}\\ \vspace{0.0cm}
\includegraphics[scale=0.51]{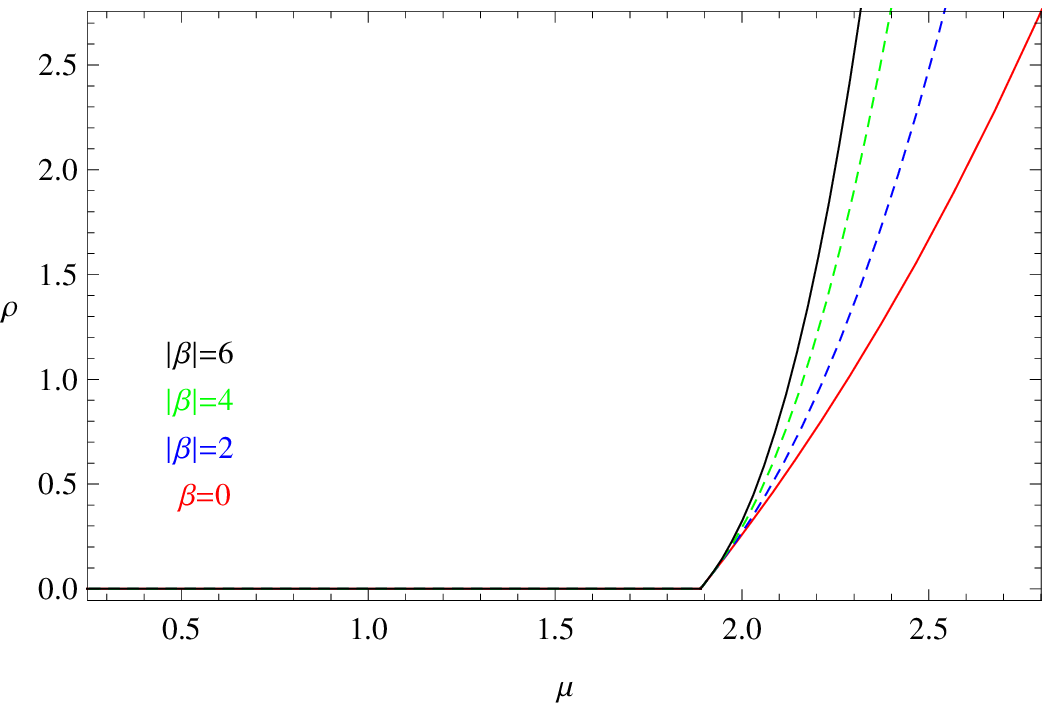}\hspace{0.2cm}%
\includegraphics[scale=0.51]{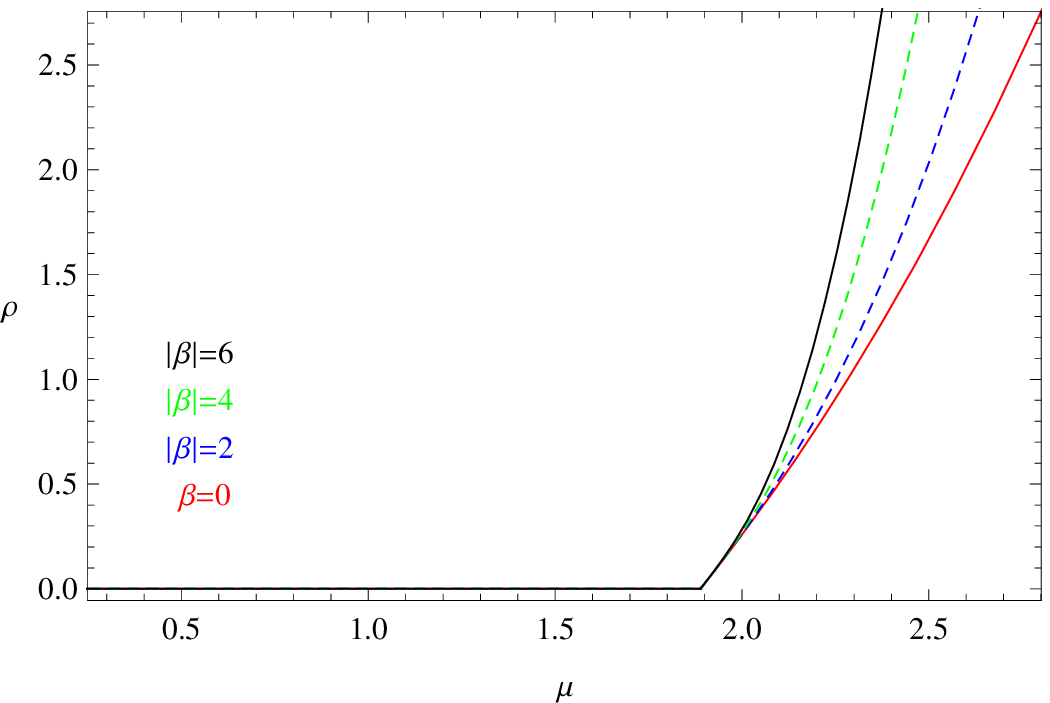}\hspace{0.2cm}%
\includegraphics[scale=0.51]{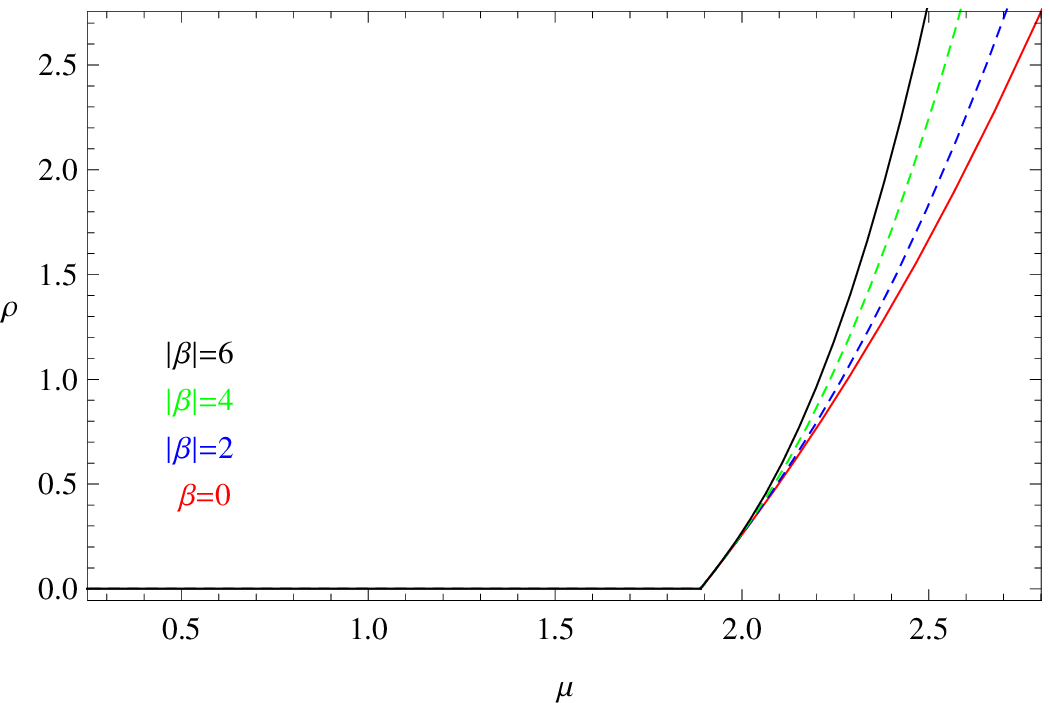}\\ \vspace{0.0cm}
\caption{\label{SolitonNLEZ375} (color online) The condensate of the operator $\cal O_{+}$ and charge density $\rho$ with the ENE (left two panels), BINE (middle two panels) and LNE (right two panels) as a function of the chemical potential $\mu$ for different nonlinearity parameters $\beta$ with the mass of the scalar field $m^{2}L^{2}=-15/4$ for the holographic superconductor and insulator model. In each panel, the four lines from bottom to top correspond to increasing absolute value of the nonlinearity parameter, i.e., $|\beta|=0$ (red), $2$ (blue), $4$ (green) and $6$ (black) respectively.}
\end{figure}

In Figs. \ref{SolitonNLEF375} and \ref{SolitonNLEZ375} we plot the condensate of the operator ${\cal O}_{i}$ ($i=-$ or $i=+$) and charge density $\rho$ with the ENE (left two panels), BINE (middle two panels) and LNE (right two panels) as a function of the chemical potential $\mu$ for different nonlinearity parameters $\beta$ with the mass of the scalar field $m^{2}L^{2}=-15/4$ for the holographic superconductor and insulator model. For the fixed $\beta$ in the each type of the nonlinear electrodynamics considered here, we observe that the system is described by the AdS soliton solution itself when $\mu$ is small, which can be interpreted as the insulator phase \cite{Nishioka-Ryu-Takayanagi}. But when $\mu\rightarrow\mu_{ci}$, there is a phase transition and the AdS soliton reaches the superconductor (or superfluid) phase for larger $\mu$. It is interesting to note that the critical chemical potential $\mu_{ci}$ for the scalar operator ${\cal O}_{i}$ is independent of the nonlinearity parameter $\beta$, i.e.,
\begin{eqnarray}
\mu_{c-}=0.836~~{\rm and}~~\mu_{c+}=1.888,\quad {\rm for}~~m^{2}L^{2}=-15/4~~{\rm and}~~\forall\beta,
\label{SolitonNLECCP}
\end{eqnarray}
which is in strong contrast to the effect of $\beta$ on the critical temperature $T_{c}$ of the holographic superconducting models with the nonlinear electrodynamics. As a matter of fact, the critical chemical potential $\mu_{ci}$ of the holographic superconductor/insulator transitions with the nonlinear electrodynamics is independent of the nonlinearity parameters $\beta$ but depends on the mass of
the scalar field, which is different from the holographic superconductor/insulator models in Gauss-Bonnet gravity \cite{Pan-Wang,Pan-Jing-Wang-Soliton}.

\subsection{Analytical study on the superconductor/insulator phase transition}

From the numerical investigation, we have found that the nonlinearity parameter $\beta$ does not have any effect on the critical chemical potentials $\mu_{c-}$ and $\mu_{c+}$ for the fixed mass of the scalar field. In this section we will apply the Sturm-Liouville method \cite{Siopsis,Li-Cai-Zhang,Cai-Li-Zhang,CaiSoliton,SiopsisBF,HGST1,PanJWCh,ChenJingPan,WangLiuCaiTZ,YYBu} to analytically investigate the properties of holographic superconductor/insulator phase transition with the nonlinear electrodynamics and back up the numerical computations.

For clarity, we will rewrite the equations of motion (\ref{solitonPsi}) and (\ref{solitonPhi}) for $d=5$ into
\begin{eqnarray}
\frac{d^{2}\psi}{dr^{2}}+\left(\frac{3}{r}+\frac{1}{f}\frac{df}{dr}\right)\frac{d\psi}{dr}
+\left(\frac{\phi^2}{r^2f}-\frac{m^2}{f}\right)\psi=0\,,
\label{solitonD5Psi}
\end{eqnarray}
\begin{eqnarray}
\frac{d\mathcal{L}}{dF^{2}}\frac{d^{2}\phi}{dr^{2}}
+\left[\left(\frac{1}{r}+\frac{1}{f}\frac{df}{dr}\right)\frac{d\mathcal{L}}{dF^{2}}
+\frac{d}{dr}\left(\frac{d\mathcal{L}}{dF^{2}}\right)\right]\frac{d\phi}{dr}+\frac{\psi^2}{2f}\phi=0,
\label{solitonD5Phi}
\end{eqnarray}
which are avaliable for the three types of the nonlinear electrodynamics considered in this work.

Introducing a new variable $z=1/r$, we can express the equations of motion (\ref{solitonD5Psi}) and (\ref{solitonD5Phi}) as
\begin{eqnarray}
\frac{d^{2}\psi}{dz^{2}}+\left(\frac{1}{f}\frac{df}{dz}-\frac{1}{z}\right)\frac{d\psi}{dz}
+\left(\frac{\phi^2}{z^2f}-\frac{m^2}{z^4f}\right)\psi=0\,,
\label{solitonD5PsiZ}
\end{eqnarray}
\begin{eqnarray}
\frac{d\mathcal{L}}{dF^{2}}\frac{d^{2}\phi}{dz^{2}}
+\left[\left(\frac{1}{z}+\frac{1}{f}\frac{df}{dz}\right)\frac{d\mathcal{L}}{dF^{2}}
+\frac{d}{dz}\left(\frac{d\mathcal{L}}{dF^{2}}\right)\right]\frac{d\phi}{dz}+\frac{\psi^2}{2z^{4}f}\phi=0.
\label{solitonD5PhiZ}
\end{eqnarray}
Considering that the scalar field $\psi=0$ at the critical chemical potential $\mu_{c}$, we obtain the equation of motion for the gauge field $\phi$ near the critical point
\begin{eqnarray}
\frac{d\mathcal{L}}{dF^{2}}\frac{d^{2}\phi}{dz^{2}}
+\left[\left(\frac{1}{z}+\frac{1}{f}\frac{df}{dz}\right)\frac{d\mathcal{L}}{dF^{2}}
+\frac{d}{dz}\left(\frac{d\mathcal{L}}{dF^{2}}\right)\right]\frac{d\phi}{dz}=0.
\label{NESWPhiCritical}
\end{eqnarray}
Obviously, the physical solution of $\phi$ to Eq. (\ref{NESWPhiCritical}) is $\phi(z)=\mu$ when $\mu<\mu_{c}$. Thus, from Eq. (\ref{solitonD5PsiZ}) we have the master equation which will be used to calculate the critical chemical potential $\mu_{c}$ in the Sturm-Liouville method
\begin{eqnarray}
\frac{d^{2}\psi}{dz^{2}}+\left(\frac{1}{f}\frac{df}{dz}-\frac{1}{z}\right)\frac{d\psi}{dz}
+\left(\frac{\mu^2}{z^2f}-\frac{m^2}{z^4f}\right)\psi=0.
\label{NESWCriticalPsi}
\end{eqnarray}
Since the nonlinearity parameter $\beta$ is absent in Eq. (\ref{NESWCriticalPsi}), we can conclude that, for the fixed mass of the scalar field, the critical chemical potential $\mu_{c}$ is independent of the explicit form of the nonlinear electrodynamics, which may be a quite general feature for the holographic superconductor and insulator model.

Working on Eq. (\ref{NESWCriticalPsi}), we can understand the dependence of the critical chemical potential on the mass of the scalar field analytically. Near the critical point, we introduce a trial function $F(z)$ into the asymptotical form of $\psi$ near the boundary $z=0$ like
\begin{eqnarray}\label{PhiFz}
\psi(z)\sim \langle{\cal O}_{i}\rangle z^{\lambda_i}F(z),
\end{eqnarray}
with $\lambda_\pm=2\pm\sqrt{4+m^{2}L^2}$. From the boundary conditions of $\psi$, we can obtain the boundary condition $F(0)=1$. For simplicity, we can set $F'(0)=0$ just as in \cite{Cai-Li-Zhang}. Thus, we can reach the equation of motion for $F(z)$
\begin{eqnarray}\label{NewFzmotion}
\frac{d}{dz}\left(M\frac{dF}{dz}\right)+M\left(P+\mu^{2}Q\right)F=0,
\end{eqnarray}
with
\begin{eqnarray}
P=\frac{\lambda_{i}(\lambda_{i}-1)}{z^{2}}+\frac{\lambda_{i}}{z}\left(\frac{1}{f}\frac{df}{dz}-\frac{1}{z}\right)
-\frac{m^{2}}{z^{4}f},~~Q=\frac{1}{z^{2}f},~~M(z)=z^{2\lambda_{i}-3}(z^{4}-1).
\end{eqnarray}
Following the Sturm-Liouville eigenvalue problem \cite{Gelfand-Fomin}, we can obtain the critical chemical potential by minimizing the following expression
\begin{eqnarray}\label{eigenvalue}
&\mu^{2}&=\frac{\int^{1}_{0}M\left[(dF/dz)^{2}-PF^{2}\right]dz}{\int^{1}_{0}MQF^{2}dz}
\nonumber\\
&&=\frac{(\lambda_{i}-2)(\lambda_{i}^{2}-1)(4+2\lambda_{i}+m^{2})\alpha^{2}
-2\lambda_{i}(\lambda_{i}-2)[(1+\lambda_{i})m^{2}+2\lambda_{i}(2+\lambda_{i})]\alpha
+\lambda_{i}(\lambda_{i}^{2}-1)(2\lambda_{i}+m^{2})}{(\lambda_{i}-2)[\lambda_{i}(\lambda_{i}-1)\alpha^{2}
-2(\lambda_{i}^{2}-1)\alpha+\lambda_{i}(\lambda_{i}+1)]},\nonumber\\
\end{eqnarray}
where we have chosen the simplest form of $F(z)$, i.e., $F(z)=1-\alpha z^{2}$ with a constant $\alpha$ in the following calculation.

For clarity, we will only discuss the case of $m^{2}L^{2}=-15/4$ and
one can easily extend the study to the cases for different values of
the mass of scalar field. From Eq. (\ref{eigenvalue}), we can obtain
the minimum eigenvalues of $\mu_{min}^{2}$ and the corresponding
values of $\alpha$ for the fixed mass of scalar field
$m^{2}L^{2}=-15/4$, i.e., $\mu_{min}^{2}=0.700$ and $\alpha=0.230$
for $i=-$ and $\mu_{min}^{2}=3.574$ and $\alpha=0.330$ for $i=+$.
Thus, we analytically get the critical chemical potential
$\mu_{c}=\mu_{min}$ \cite{Cai-Li-Zhang}
\begin{eqnarray}
\mu_{c-}=0.837~~{\rm and}~~\mu_{c+}=1.890,\quad {\rm for}~~m^{2}L^{2}=-15/4~~{\rm and}~~\forall\beta,
\label{AnalyticNLECCP}
\end{eqnarray}
which agrees well with the numerical result presented in Eq. (\ref{SolitonNLECCP}). Obviously, our analytic results back up the numerical findings in the holographic superconductor/insulator transitions with the nonlinear electrodynamics and show that the nonlinearity parameter $\beta$ really has no effect on the critical chemical potentials $\mu_{c}$ of the phase transitions.

\section{Conclusions}

In order to investigate systematically the effect of the nonlinear correction to the usual Maxwell electrodynamics, we have introduced the holographic dual models with three kinds of typical Born-Infeld-like nonlinear electrodynamics both in the backgrounds of AdS black hole and AdS soliton. Considering that these nonlinear generalizations essentially imply the higher derivative corrections of the gauge fields, this study may help to understand the influences of the $1/N$ or $1/\lambda$ corrections on the holographic dual models. Comparing with the Born-Infeld nonlinear electrodynamics (BINE) and Logarithmic form of nonlinear electrodynamics (LNE), in the black hole background, we found the Exponential form of nonlinear electrodynamics (ENE) has stronger effect on the condensation formation and conductivity for the holographic superconductors. Furthermore, similar to the curvature correction, we observed that for all three types of the nonlinear electrodynamics considered here the higher nonlinear electrodynamics correction term can make the condensation harder to form and result in the larger deviations from the universal value $\omega_g/T_c\approx 8$ for the gap frequency. Thus, we argued that the nonlinear electrodynamics correction and the Gauss-Bonnet correction share some similar features for the holographic superconductor system if the nonlinear electrodynamics takes the form of this work. However, the story is completely different if we study the holographic superconductor/insulator transitions with the nonlinear electrodynamics. In contrast to the curvature correction effect, we found that in the AdS soliton background the critical chemical potentials are independent of the explicit form of the nonlinear electrodynamics, i.e., the ENE, BINE and LNE correction terms do not have any effect on the critical chemical potentials. We confirmed our numerical result by using the Sturm-Liouville analytic method and concluded that the nonlinear correction to the usual Maxwell electrodynamics will not affect the properties of the holographic superconductor/insulator phase transitions, which may be a quite general feature for the s-wave holographic superconductor and insulator system.

\begin{acknowledgments}

This work was supported by the National Natural Science Foundation of China under Grant Nos. 11275066 and 11175065; the National Basic Research of China under Grant No. 2010CB833004; PCSIRT under Grant No. IRT0964; NCET under Grant No. 10-0165; Hunan Provincial Natural Science Foundation of China under Grant Nos. 12JJ4007 and 11JJ7001; and the Construct Program of the National Key Discipline.

\end{acknowledgments}

\end{document}